\documentclass[11pt]{article}
\usepackage{graphicx,epsfig}
\usepackage{color}
\usepackage{amsmath,amsthm,amssymb}  
\usepackage{mathrsfs}

\usepackage[english,francais]{babel}

\usepackage[latin1]{inputenc}  

\def\eqdef{\stackrel{\mbox{\tiny def}}{=}}     
\newcommand{\ket}[1]{|\kern.3ex#1\kern.3ex\rangle}
\newcommand{\bra}[1]{\langle\kern.3ex #1 \kern.3ex|}
\newcommand{\scalar}[2]{\langle\kern.3ex #1 \kern.3ex|\kern.3ex#2\kern.3ex\rangle}

\newcommand{\EXP}[1]{{\mbox{\large e}}^{#1}}         

\newcommand{\tr}[1]{\mathop{\mathrm{Tr}}\nolimits\left\{ #1 \right\}}  
  
\renewcommand{\max}[2]{\mathop{\mathrm{max}}\nolimits\left( #1 , #2\right)}

\def\I{{\rm i}}                  
\def\D{{\rm d}}                  
\def\Dc{{\rm D}}                 

\newcommand{\deriv}[2]{\frac{\mathrm{d}#1}{\mathrm{d}#2}}
\newcommand{\derivp}[2]{\frac{\partial #1}{\partial #2}}

\newcommand\ab{{\alpha\beta}}
\newcommand\ba{{\beta\alpha}}

\newcommand\tc{\widetilde{\mathcal{C}}}

\begin{document}

\selectlanguage{english}

\title{$\zeta$-regularised spectral determinants on metric graphs} 

\author{Christophe Texier$^{(a,b)}$}

\date{June 10, 2010}

\maketitle

\hspace{1cm}
$^{(a)}$
\begin{minipage}[t]{13cm}
{\small
Univ. Paris Sud ; CNRS ; LPTMS, UMR 8626, 
B\^at. 100, F-91405 Orsay, France.

}
\end{minipage}

\vspace{0.25cm}

\hspace{1cm}
$^{(b)}$
\begin{minipage}[t]{13cm}
{\small
Univ. Paris Sud ; CNRS ; LPS, UMR 8502, 
B\^at. 510, F-91405 Orsay, France.

}
\end{minipage}

\begin{abstract}
  Several general results for the spectral determinant 
  of the Schr\"o\-dinger operator on metric graphs are reviewed. 
  Then, a simple derivation for the $\zeta$-regularised spectral
  determinant is proposed, based on the Roth trace formula. 
  Two types of boundary conditions are studied~: functions continuous
  at the vertices and functions whose derivative is continuous at the
  vertices. 
  The $\zeta$-regularised spectral determinant of the Schr\"o\-dinger
  operator acting on functions with the most general boundary
  conditions is conjectured in conclusion.
  The relation to the Ihara, Bass and Bartholdi formulae
  obtained for combinatorial graphs is also discussed.
\end{abstract}

\noindent
PACS numbers~: 02.70.Hm~; 02.10.Ox 





\section{Introduction}

Spectral determinants (i.e. $\zeta$-functions or $L$-functions) and trace
formulae in graphs have attracted a lot of interest in the 
mathematical literature \cite{Iha66,Rot83,Rot83a,Has89,Has90,Bas92,For93,StaTer96,Bar99,Che99,Fri06,KosSch07,HarKir09} and in the
physical literature as well \cite{Pas98,PasMon99,BerKea99,KotSmi99,AkkComDesMonTex00,Des01,Des02,ComDesMaj02,KeaMarWin03,FerAngRowGueBouTexMonMai04,TexMon05,ComDesTex05}.
In this article we will consider the case of the
Laplace $-\Delta$ or the Schr\"odinger operator $-\Delta+V(x)$ on a
metric graph. 
We denote by $\{E_n\}$ the spectrum of the operator $-\Delta+V(x)$,
and introduce the spectral determinant by the formal definition
\begin{equation}
  \label{eq:FormalDef}
  S(\gamma)=\det[\gamma-\Delta+V(x)]=\prod_n(\gamma+E_n) 
  \:,
\end{equation}
where
$\gamma$ is some ``spectral parameter''.
This object was shown to be a very convenient quantity in order to
study several questions in metric graphs, like various properties of
the Brownian motion
\cite{Des01,Des02,ComDesMaj02,TexMon05,ComDesTex05} or transport and
magnetic properties of networks of metallic wires 
\cite{Pas98,PasMon99,AkkComDesMonTex00,FerAngRowGueBouTexMonMai04}. 

Because the operator $-\Delta+V(x)$ acts in a space of
infinite dimension, the computation of the spectral determinant
\eqref{eq:FormalDef} requires in practice some regularisation.
Starting from the trace of the Green function, i.e. the Laplace
transform of the partition function $Z(t)=\sum_n\EXP{-tE_n}$, 
\begin{equation}
  G(\gamma)
  =\sum_n\frac1{\gamma+E_n}=\int_0^\infty\D{}t\,Z(t)\,\EXP{-\gamma{}t}
  \:,
\end{equation}
furnishes a first possible regularisation, used in
Refs.~\cite{Pas98,PasMon99,AkkComDesMonTex00,Des00,Des01,TexMon05,ComDesTex05,Tex08}.
Performing some {\it integration} with respect
to the spectral parameter we obtain
\begin{equation}
  \label{eq:SGF}
  S^\mathrm{GF}(\gamma) \eqdef 
  \exp\int_{\gamma_0}^\gamma\D\gamma'\,
  G(\gamma')
  = S(\gamma) / S(\gamma_0)
\end{equation}
For example, consider the Laplace operator on a line $[0,L]$
with Neumann boundary conditions. The spectrum is in this case 
$E_n=(n\pi/L)^2$ with $n\in\mathbb{N}$ and we obtain
$S^\mathrm{GF}(\gamma)=\frac{\sqrt\gamma\sinh\sqrt\gamma{}L}{\sqrt\gamma_0\sinh\sqrt\gamma_0{}L}$.
This procedure however leaves some arbitrary in the choice of the
parameter $\gamma_0$.

Another well known regularisation for determinants is the
$\zeta$-regularisation whose starting point is to introduce the
$\zeta$-function 
\begin{equation}
  \zeta(s,\gamma) \eqdef \sum_n{(\gamma+E_n)^{-s}}
  = \frac1{\Gamma(s)}\int_0^\infty\D{}t\,t^{s-1}\,
  \EXP{-\gamma t}\,Z(t)
  \:,
\end{equation}
that we have expressed as a Mellin transform of the partition
function, for convenience for the following discussion. 
Note that $\zeta(1,\gamma)=G(\gamma)$. The $\zeta$-regularised
determinant is then related to a {\it derivative} of the
$\zeta$-function~:  
\begin{equation}
  S^\zeta(\gamma) \eqdef \exp-\deriv{\zeta}{s}(0,\gamma)
  \:.
\end{equation}
In this case no arbitrary is left within the calculation of
the determinant.
Coming back to the simple example considered above of the Laplace
operator on a line $[0,L]$ 
with Neumann boundary conditions, we get in this case
$S^\zeta(\gamma)=2\sqrt\gamma\sinh\sqrt\gamma{}L$ (derived below).

An important remark is that various regularisations only
differ by some $\gamma$-independent prefactors. This explains why the
choice of the regularisation has no consequence on the properties
studied in
Refs.~\cite{Pas98,PasMon99,AkkComDesMonTex00,Des01,Des02,ComDesMaj02,FerAngRowGueBouTexMonMai04,TexMon05,ComDesTex05}
since they are always related to derivatives of
$\ln{}S(\gamma)$, with respect to the spectral parameter $\gamma$ or
other parameters. 
In a recent work \cite{Tex08} some procedure was proposed in
order to construct the spectral determinant of a graph by combinations
of the determinants of subgraphs. In
this case it is crucial to define precisely the prefactor of the
spectral determinant.

In a recent paper \cite{Fri06}, Friedlander has derived the
$\zeta$-regularised determinant for the Schr\"o\-din\-ger operator on
a metric graph.  
One should also mention that the analysis of the regularised
determinant of the Laplace operator for
general boundary conditions, $\det'(-\Delta)$ with the prime
indicating exclusion of zero mode contribution is there is some, was
the subject of the very recent work \cite{HarKir09}. 
It is the purpose of the present article to propose another derivation
of the $\zeta$-regularised determinant for the Schr\"o\-din\-ger
operator $\det[\gamma-\Delta+V(x)]$. 
Our approach is based on the Roth trace formula \cite{Rot83} and will
allow simple extension of Friedlander's result to other
choices of boundary conditions at the vertices. 

In the next section we set notations.
In section \ref{sec:Desbois} we mostly recall some known general results
obtained by Desbois in Ref.~\cite{Des01} and needed for the following
sections. 
The section \ref{sec:cbc} focuses on the case of functions continuous
at the vertices and section \ref{sec:cdbc} on the case of functions
with derivative continuous at vertices.
The $\zeta$-regularisation of the spectral determinant is provided in
section~\ref{sec:zr}.

\section{Metric graphs and Laplace operator}

A graph is a set of $V$ vertices (here labelled with greek letters
$\alpha$, $\beta$,...) linked by $B$ bonds (denoted as $(\ab)$,...).
Each bond $(\ab)$ is associated to two arcs (oriented bonds) $\ab$ and
$\ba$  (arc will be also labelled with roman indices $a$, $b$,....). 
For the arc $a=\alpha\beta$, the reversed arc is denoted
$\bar{a}=\beta\alpha$. 
We introduce the adjacency matrix~: $a_\ab=1$ if $\alpha$ is linked to
$\beta$ by a bond and $a_\ab=0$ otherwise.
We denote by $m_\alpha=\sum_\beta{}a_\ab$ the coordination number (valency) of
the vertex.
The graph is said to be a {\it metric graph}
(or a quantum graph) if each bond is identified with a finite
interval $(0,l_\ab)\in\mathbb{R}$, where $l_\ab$ designates the length
of the bond linking the vertices $\alpha$ and $\beta$.
In this case we may consider a scalar function $\psi(x)$,
characterised by a set of $B$ components $\psi_\ab(x_\ab)$ with
$x_\ab\in(0,l_\ab)$. 
The component is labelled by the arc in order to specify the direction
along which the coordinate is measured [this redundancy of the
notation implies the obvious relation 
$\psi_\ab(x_\ab)=\psi_\ba(x_\ba)=\psi_\ba(l_\ab-x_\ab)$]. 

Having introduced scalar functions living on the graph we may define
the action of the Laplace operator $\Delta$ on such functions.
Along a wire it acts as the usual second derivative
$(\Delta\psi)_\ab(x)=\deriv{^2}{x^2}\psi_\ab(x)$. 
%
At the vertices, the set of functions on which $\Delta$ acts must
satisfy some boundary conditions in order to ensure self adjointness of
the operator.
Let us denote by $\psi(0)$ and $\psi'(0)$ the vectors of size $2B$
gathering the 
values taken by the components and the derivatives at the vertices
$\psi_\ab(0)$ and $\psi'_\ab(0)$, respectively. The most general
boundary conditions ensuring self adjointness of $\Delta$ are of the form
\begin{equation}
  \label{eq:bc}
  C\, \psi(0) + D\, \psi'(0) = 0
  \:,
\end{equation}
where the $2B\times2B$ matrices $C$ and $D$
satisfy ({\it i}) $CD^\dagger=DC^\dagger$. ({\it ii})
The $2B\times4B$ matrix $(C,D)$ has maximal rank (equal to $2B$)
\cite{KosSch99}. 
Note that, characterising what arc is connected to what other arc, these two
matrices encode all the information on the topology of the graph.

\section{General results for the spectral determinant}
\label{sec:Desbois}

In this section we mostly recall some results obtained by Desbois
\cite{Des01} for the spectral determinant of the Schr\"odinger
operator for general boundary conditions.
We compute the spectral determinant by constructing the 
Green function 
$\mathcal{G}(x,y)\eqdef\bra{x}\frac1{\gamma-\Dc_x^2+V(x)}\ket{y}$,
where $\Dc_x\eqdef\deriv{}{x}-\I{}A(x)$ is the covariant derivative.
Note that the replacement of $\Delta$ by $\Dc_x^2$ is motivated by
physical considerations \cite{PasMon98,AkkComDesMonTex00,TexMon01}. 
It might also be
useful in order to study winding properties of Brownian curves in the
graph \cite{TexMon05,ComDesTex05}. It will affect calculations in a
very simple manner through the introduction of additional (magnetic)
phases. 
In the following it will be understood that the $2B$-vector $\psi'(0)$
of Eq.~\eqref{eq:bc} gathers the covariant derivatives.

\begin{figure}[!ht]
  \centering
  \includegraphics[scale=0.9]{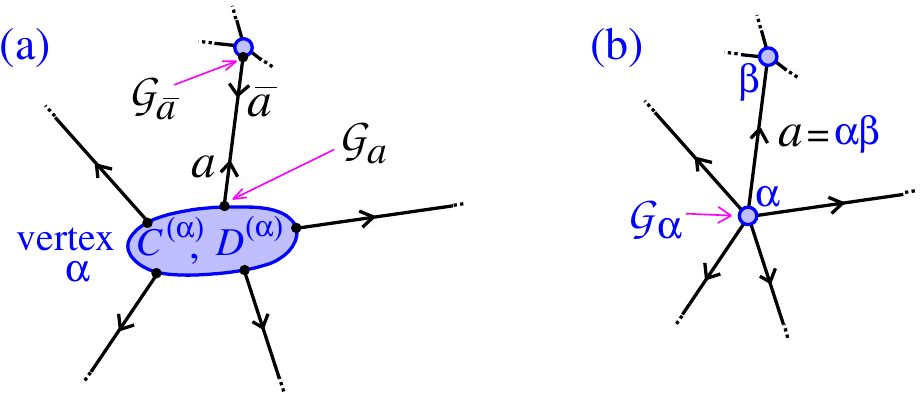}
  \caption{{\it The vertex $\alpha$.} 
    (a) {\it For general boundary conditions,
    matching of the Green function at the vertex is
    described through two 
    matrices $m_\alpha\times m_\alpha$ denoted $C^{(\alpha)}$ and
    $D^{(\alpha)}$ (the blocks of the matrices $C$ and $D$). One must
    assign a value of the Green function $\mathcal{G}_a$ to each
    arc $a$.}
    (b) {\it For continuous boundary condition at the vertex, the
      matrices $C^{(\alpha)}$ and $D^{(\alpha)}$ take the form
    \eqref{eq:CDcont}. In this case one can assign to each vertex $\alpha$ a
    single ``vertex variable'' $\mathcal{G}_\alpha$.}}
  \label{fig:vertex}
\end{figure}

\subsection{Arc determinant (1)}

Let us consider two points $x$ and $y$ belonging to the two
arcs $a$ and $b$, respectively.
On the arc $a$ we use the set of independent solutions $f_a$ and
$f_{\bar a}$ of the differential equation
$[\gamma-\deriv{^2}{x^2}+V_a(x)]f(x)=0$ such that $f_a(0)=1$ and
$f_a(l_a)=0$ (see appendix~\ref{sec:bondscatt}). 
The Green function depends on two coordinates $x$ and $y$ and
therefore on must specify by two indices (here $a$ and $b$) to which
arcs they belong to~:
\begin{align}
  \label{eq:Gab}
  \mathcal{G}_{a,b}(x_a,y_b) = & 
    \mathcal{G}_a \, f_a(x_a)\,\EXP{\I A_ax_a} 
  + \mathcal{G}_{\bar a} \, f_{\bar a}(x_{\bar a})\,\EXP{\I A_{\bar a}x_{\bar a}}
  \nonumber\\
  & + \delta_{(a),(b)}
  \frac{\EXP{\I A_a(x_a-y_{a})} }
       {{W}_a}\:
  f_a(\max{x_a}{y_a})\:
  f_{\bar a}(\max{x_{\bar a}}{y_{\bar a}})   
  \:,
\end{align}
where $\mathcal{G}_a$ and $\mathcal{G}_{\bar a}$ denote the values of
the Green function at the two ends of the bond $(a)$ (see
figure~\ref{fig:vertex}.a). 
We have introduce the notation
$\delta_{(a),(b)}\eqdef\delta_{a,b}+\delta_{a,\bar b}$.
Note that the dependence of $\mathcal{G}_a$ and $\mathcal{G}_{\bar a}$
in the coordinate $y$ is implicit.
$A_a=A(x)$ for $x\in(a)$ is the (constant) vector potential on the
bond~\footnote{
  In one dimension, a dependence of the vector potential in the
  coordinate may always be removed by a convenient gauge transformation.
} and
$\theta_a=A_al_a$ the magnetic flux along the wire.
${W}_a=-f'_a(l_a)$ is the Wronski determinant of the two linearly
independent solutions, Eq.~\eqref{eq:Wronskif}.
All the values of the Green function at the vertices
$\mathcal{G}_{a,b}(0,y_b)=\mathcal{G}_a$ (figure~\ref{fig:vertex}.a)
are gathered in the vector 
$\mathcal{G}(0)$ of size $2B$ and all covariant derivatives 
$(\Dc_x\mathcal{G})_{a,b}(0,y_b)$ in the vector $\mathcal{G}'(0)$. We obtain
\begin{equation}
  \mathcal{G}'(0) = M\,\mathcal{G}(0) 
  +\left(
    \begin{array}{cccccc}
      \vdots \\ 0 \\ f_b(y_b)\,\EXP{-\I A_by_b}\\
      f_{\bar b}(y_{\bar b})\,\EXP{-\I A_{\bar b}y_{\bar b}} \\ 0 \\ \vdots
    \end{array}
  \right)   
    \begin{array}{cccccc}
      \phantom{\vdots} \\ \phantom{0} \\ \leftarrow b\\ 
      \leftarrow \bar b \\ \phantom{0} \\ \phantom{\vdots} 
    \end{array}
\end{equation}
where the matrix $M$ couples an arc $a$ to itself and to the reversed
arc $\bar{a}$~: 
\begin{equation}
  \label{eq:defNJean}
  M_{a,b} \eqdef \delta_{a,b}\, f'_a(0)
  - \delta_{a,\bar b}\, f'_{\bar a}(l_a)\,\EXP{\I\theta_{\bar a}}
  \:.
\end{equation}
Noticing that  $f'_{\bar a}(l_a)=f'_{a}(l_a)\in\mathbb{R}$ (see
appendix~\ref{sec:bondscatt}) and $\theta_{\bar a}=-\theta_{a}$
shows that this matrix is Hermitian.
Imposing boundary conditions at vertices,
$C\,\mathcal{G}(0)+D\,\mathcal{G}'(0)=0$, we obtain all values
$\mathcal{G}_a$, that could be reinjected into Eq.~\eqref{eq:Gab}. 
The Green function at coinciding points reads~:
\begin{align}
  \label{eq:fctGgbc}
  \mathcal{G}_{a,a}(x_a,x_a) &= - \left[(C+DM)^{-1}D\right]_{aa}\,f_a^2  
 - \left[(C+DM)^{-1}D\right]_{a\bar a}\,
   f_a f_{\bar a}\EXP{\I\theta_{a}}
\nonumber \\ &
 - \left[(C+DM)^{-1}D\right]_{\bar aa}\,
   f_a f_{\bar a}\EXP{\I\theta_{\bar a}}
 - \left[(C+DM)^{-1}D\right]_{\bar a\bar a}\,
   f_{\bar a}^2
 + \frac1{{W}_a}f_a f_{\bar a}
\end{align}
Integration over the Graph may be decomposed as integration over the
bonds~:
$\int_\mathrm{Graph}\D{}x\,\mathcal{G}(x,x)=\sum_{(a)}\int_{0}^{l_a}\D{}x_a\,\mathcal{G}_{a,a}(x_a,x_a)$.
The integration of product of functions $f_a$ and $f_{\bar{a}}$ is
given by eqs.~(\ref{eq:RelDes00a},\ref{eq:RelDes00b}), hence
\begin{align}
  \label{eq:fctGgbc2}
  G(\gamma)= \int_\mathrm{Graph}\D{}x\, \mathcal{G}(x,x) 
  = \tr{ (C+DM)^{-1}D\,\partial_\gamma M } -
  \partial_\gamma\sum_{(a)}\ln[-f'_{a}(l_a)]
  \:,
\end{align}
where the trace runs over the $2B$ arc indices.
Up to now, nothing has be assumed on the nature of the
boundary conditions. An expansion of the trace provides a
trace formula for $\int_\mathrm{Graph}\D{}x\,\mathcal{G}(x,x)$, where
the expansion is interpreted as sum of contributions over cycles
(orbits) in the graph.

We now suppose that the matrices $C$ and $D$ {\it do not
depend on the spectral parameter} $\gamma$. Performing and integration
$\ln S(\gamma)=\int^\gamma\D\gamma'\,G(\gamma')$, we obtain the first
important result of Desbois~:
\begin{equation}
  \label{eq:Sgbc1JD}
  \boxed{
    S(\gamma) = (-1)^V \prod_{(a)}\frac{-1}{f'_a(l_a)}\:\det(C+DM)
  }
\end{equation}
where the product runs over the $B$ bonds.
This provides a general expression of the spectral determinant in terms
of $2B\times2B$-matrices coupling arcs.
We repeat that the matrix $M$ couples reversed arcs and 
contains local information related to the potential on each bound. On
the other hand, the matrices $C$ and $D$, characterising which arcs
arrive and issue from the vertices, encode the information on the
topology of the graph.
Note that the structure $\det(C+DM)$ was also obtained in
\cite{HarKir09} in the absence of the potential 
where the determinant $\det'(-\Delta)$ was analysed (the prime
indicates exclusion of zero mode).
When $V(x)=0$ the matrix $M$ takes the form
$M_{a,b}=-\sqrt\gamma\big(\delta_{a,b}\, \coth\sqrt\gamma l_a
-\delta_{a,\bar b}\,\frac1{\sinh\sqrt\gamma l_a}\,\EXP{\I\theta_{\bar a}}\big)$.

The prefactor in Eq.~\eqref{eq:Sgbc1JD} has
been chosen arbitrarily and for convenience for the following. We will
come back to the question of the prefactor in the section~\ref{sec:zr}.

\paragraph{Limit \mathversion{bold}$\gamma\to\infty$\mathversion{normal}}
For large positive spectral parameter we can ignore the potential,
therefore $f_a'(0)\simeq-\sqrt\gamma$ and
$f_a'(l_a)\simeq-\sqrt\gamma\EXP{-\sqrt\gamma\,l_a}$. 
Hence the matrix $M$ becomes diagonal~: 
$M_{a,b}\simeq-\delta_{a,b}\sqrt\gamma$ and we obtain
\begin{equation}
  S(\gamma) \underset{\gamma\to+\infty}{\simeq}(-1)^V
  2^{-B}\gamma^{-B/2}\EXP{\sqrt\gamma\,\mathcal{L}}\:
  \det(C-\sqrt\gamma\,D)  
\end{equation}
where $\mathcal{L}=\sum_{(a)}l_a$ is the ``total length'' of the graph.

\paragraph{Example}
Let us consider a ring of perimeter $L$ pierced by a magnetic flux $\theta$
(a graph in which the number of vertices can be
reduced to $V=1$ with $B=1$ bond).
We denote by $1$ and $\bar1$ the two reversed arcs.
Let us consider boundary conditions described by the matrices given
below by
Eq.~\eqref{eq:CDcont}. We obtain 
\begin{equation}
  S(\gamma) = 
  \frac{\lambda - f_1'(0) - f_{\bar 1}'(0) + 2 f_1'(L)\,\cos\theta }{-f_1'(L)}  
  \:.
\end{equation}
It is interesting to point that this formula has found a practical
physical application for a potential of the form $V(x)\propto{}x(L-x)$
in the context of the study of decoherence by electron-electron
interaction in a phase coherent metallic ring~\cite{TexMon05b} (hence
$f_1(x)$ is a Hermite function and $f_{\bar 1}(x)=f_1(L-x)$ due to the
property $V(x)=V(L-x)$). 

The case $V(x)=0$ is analysed in section~\ref{sec:zr}.


\subsection{Arc determinant (2)}
The above derivation has privileged the use of variables $\mathcal{G}_a$
giving the value of the Green function at the vertices. Another
natural choice is to deal with (covariant) derivatives of the Green
function at the vertices. We write
\begin{align}
  \label{eq:GabConDer}
   \mathcal{G}_{a,b}(x_a,y_b) =& 
   \mathcal{G}'_a \, g_a(x_a)\,\EXP{\I A_ax_a} 
  + \mathcal{G}'_{\bar a} \, g_{\bar a}(x_{\bar a})\,\EXP{\I A_{\bar a}x_{\bar a}}
  \nonumber\\
  &+\delta_{(a),(b)}
  \frac{\EXP{\I A_a(x_a-y_{a})} }
        {{W}^g_a}\:
  g_a(\max{x_a}{y_a})\:
  g_{\bar a}(\max{x_{\bar a}}{y_{\bar a}})   
\end{align}
where the functions $g_a$ and $g_{\bar a}$ are solutions of the
Schr\"odinger equation on the bond $(a)$ for fixed values of the
{\it derivative} at the boundaries, $g'_a(0)=1$ and $g'_a(l_a)=0$ (see
appendix~\ref{sec:bondscatt}). 
Following the lines of the previous subsection, we obtain 
\begin{equation}
  \mathcal{G}(0) = N\,\mathcal{G}'(0) 
  -\left(
    \begin{array}{cccccc}
      \vdots \\ 0 \\ g_b(y_b)\,\EXP{-\I A_by_b}\\
      g_{\bar b}(y_{\bar b})\,\EXP{-\I A_{\bar b}y_{\bar b}} \\ 0 \\ \vdots
    \end{array}
  \right)   
    \begin{array}{cccccc}
      \phantom{\vdots} \\ \phantom{0} \\ \leftarrow b\\ 
      \leftarrow \bar b \\ \phantom{0} \\ \phantom{\vdots} 
    \end{array}
\end{equation}
where we have introduced the arc matrix 
\begin{equation}
  N_{a,b} \eqdef \delta_{a,b}\, g_a(0)
  + \delta_{a,\bar b}\, g_{\bar a}(l_a)\,\EXP{\I\theta_{\bar a}}
  \:.
\end{equation}
Imposing the boundary conditions at the vertices we obtain the
values $\mathcal{G}'_a$.
Reinjecting these expressions in Eq.~\eqref{eq:GabConDer}, integration over
the coordinate and summation over bonds give another general trace formula
for the Green function
\begin{align}
  G(\gamma)= \int_\mathrm{Graph}\D{}x\, \mathcal{G}(x,x) 
  = \tr{ (CN+D)^{-1}C\,\partial_\gamma N } -
  \partial_\gamma\sum_{(a)}\ln[-g_{a}(l_a)]
  \:.
\end{align}
For $\gamma$-independent matrices $C$ and $D$, 
integration over the spectral parameter leads to
\begin{equation}
  \label{eq:DualJean}
  \boxed{
    S(\gamma) = 
    (-1)^V \prod_{(a)}\frac{-1}{g_a(l_a)}\:\det(CN+D)
  }
\end{equation}
where the $\gamma$-independent prefactor was chosen in order to match
with \eqref{eq:Sgbc1JD}.
This expression can be directly related to \eqref{eq:Sgbc1JD}  by
using the relations (demonstrated in the appendix~\ref{sec:bondscatt}) 
\begin{equation}
  N=M^{-1}
  \hspace{0.5cm}\mbox{and}\hspace{0.5cm}
  \det M = \frac1{\det N}
  =\prod_{(a)}\frac{f'_a(l_a)}{g_a(l_a)}
  \:.\label{eq:6}
\end{equation}

\paragraph{Example}
We consider again the very simple case of a ring, but this time we
choose to consider boundary conditions of the type described below by
Eq.~\eqref{eq:CDcontder}. We obtain  
\begin{equation}
  S(\gamma) = 
  \frac{\mu - g_1(0) - g_{\bar 1}(0) - 2 g_1(L)\,\cos\theta }{-g_1(L)}  
  \:.
\end{equation}
In the absence of potential we obtain
\begin{equation}
  S(\gamma) = \mu \sqrt\gamma\sinh\sqrt\gamma L
  + 2(\cosh\sqrt\gamma L + \cos\theta)
  \:.
\end{equation}
The limit $\mu\to\infty$ gives the Neumann determinant of the
bond~:
$
S(\gamma)\to\mu \sqrt\gamma\sinh\sqrt\gamma L
=\mu L\gamma
\prod_{n=1}^\infty(1+(\frac{\sqrt\gamma L}{n\pi})^2)
$.

\subsection{Arc determinant (3), scattering matrices and $\zeta$-function}

The result \eqref{eq:Sgbc1JD} can be reorganised more
conveniently by introducing the arc matrix
\begin{equation}
  \label{eq:RelRN}
  R = (\sqrt\gamma + M)(\sqrt\gamma - M)^{-1}  
\end{equation}
i.e.
$M=\sqrt\gamma(R+\mathbf{1}_{2B})^{-1}(R-\mathbf{1}_{2B})=\sqrt\gamma(R-\mathbf{1}_{2B})(R+\mathbf{1}_{2B})^{-1}$,
where $\mathbf{1}_{2B}$ is the $2B\times2B$ identity matrix. 
We denote its matrix elements
\begin{equation}
  R_{a,b} = \delta_{a,b}\, r_a 
  + \delta_{a,\bar b}\, t_{\bar a}\,\EXP{\I\theta_{\bar a}}
  \:.
\end{equation}
These matrix elements have a clear physical meaning~: $r_a$ and $t_a$
are analytic continuations (to negative 
energies $E\to-\gamma$) of reflection
and transmission probability amplitudes through the potential
$V_a(x)$ (see appendix~\ref{sec:bondscatt})~: $R$ is therefore the analytic conitnuation
of the bond scattering matrix. We also introduce the analytic
continuation of the vertex scattering matrix (scattering matrix
interpretation is discussed in Refs.~\cite{KosSch99,TexMon01,TexBut03})
\begin{equation}
  \label{eq:Qmatrix}
  Q = ( -C + \sqrt\gamma\, D )^{-1} ( C + \sqrt\gamma\, D )  
  \:,
\end{equation}
(for instance we can check that $Q$ is unitary for
$\sqrt\gamma=-\I{}k$, reflecting current conservation at the vertex).
We obtain another interesting result of Desbois~\cite{Des01}~:
\begin{equation}
  \label{eq:Sgbc2JD}
  \boxed{
    S(\gamma) = (-1)^V\prod_{(a)}\frac{-1}{f'_a(l_a)}\:
    \frac{\det(C-\sqrt\gamma\, D)}{\det(\mathbf{1}_{2B}+R)}\,
    \det(\mathbf{1}_{2B}-QR)
  }
\end{equation}
Let us describe the structure of this result~: the product
over bonds $\prod_{(a)}f'_a(l_a)$ and the determinant $\det(\mathbf{1}_{2B}+R)$
involve (local) information about the potential on the bonds (recall
that $R$ couples an arc to itself and its reversed arc, only).
This is even more clear from Eq.~\eqref{eq:useful6} that leads to
\begin{equation}
  \prod_{(a)}[-f'_a(l_a)]\,\det(\mathbf{1}_{2B}+R)
  =2^B\gamma^{B/2}\prod_{(a)}t_a
  =(4\gamma)^{B/2}\big[\prod_a\,R_{a,\bar{a}}\big]^{1/2}
  \:,\label{eq:7}
\end{equation}
where the first product runs over the $B$ bonds and the last one over
the $2B$ arcs.
Now let us consider the determinant $\det(C-\sqrt\gamma\,D)$~:
organising the arcs by gathering arcs issuing from the same vertex,
the matrices $C$ and $D$ take some block diagonal structure, hence
$\det(C-\sqrt\gamma\,D)$ encodes some (local) informations about
the vertices. 
The last determinant $\det(\mathbf{1}_{2B}-QR)$ mixes informations
about potential and the connection of arcs to vertices in a nontrivial
way, and characterises the (global) information about the topology. 

This last part of the spectral determinant vanishes on the spectrum of
$-\Dc_x^2+V(x)$, i.e. for $\gamma=-E_n$. 
Because $Q$ and $R$ have the meaning of scattering matrices, the equation
$\det(\mathbf{1}_{2B}-QR)=0$ may be understood as a quantisation
condition \`a la 
Bohr-Sommerfeld.

\vspace{0.25cm}

It is well known that the determinant $\det(\mathbf{1}_{2B}-QR)$ can
be expanded in terms of {\it primitive cycle} contributions, with the
structure of a $\zeta$-function.
We recall that a cycle (a periodic orbit) is the equivalence class of
all ordered sets 
of arcs $\mathcal{C}=(a_1,a_2,\cdots,a_n)$ identical by cyclic
permutations and such that $\forall i\in\{1,\cdots,n\}$
end$(a_{i-1})$=begining$(a_{i})$ (with $a_0\equiv a_n$).
An orbit is said primitive, and denoted $\tc$, if it cannot be
decomposed as a repetition of a smaller orbit. 
We define the weight of the orbit $\mathcal{C}$ as 
$v(\mathcal{C})\,b(\mathcal{C})=(QR)_{a_1a_2}(QR)_{a_2a_3}\cdots(QR)_{a_na_1}$
where we have identified the parts related to the scattering by the vertices
$v(\mathcal{C})=Q_{a_1b_1}Q_{a_2b_2}\cdots Q_{a_nb_n}$ and the
scattering by the bonds
$b(\mathcal{C})=R_{b_1a_2}R_{b_2a_3}\cdots R_{b_na_1}$, respectively 
(note that $b_{i-1}\in\{a_{i},\,\bar a_{i}\}$).
The determinant may be rewritten as an infinite product over
primitive orbits as
\begin{equation}
  \det(\mathbf{1}_{2B}-QR) = \prod_{\tc} 
  \big( 1 - v(\tc)\,b(\tc) \big)
\end{equation}
(see for example Refs.~\cite{StaTer96,AkkComDesMonTex00} and references
therein).
This relation emphasizes that the spectral determinant may be
interpreted as a $\zeta$-function (or $L$-function) for primitive
orbits of graphs. 
$\zeta$-functions are powerful tools of particular importance, in
number theory and graph theory for example, since they play the role of
generating functions for primitive elements. The most famous
$\zeta$-function is the Riemann $\zeta$-function 
$\zeta(s)^{-1}=\prod_{\mathrm{prime}\:p}(1-p^{-s})$ giving access to the
distribution of prime numbers.

\vspace{0.25cm}

To close this section, let us emphasize that the representations
(\ref{eq:Sgbc1JD},\ref{eq:DualJean},\ref{eq:Sgbc2JD}) are remarkable
in the sense that, despite the Laplace operator acts on a space of
infinite dimension, its spectral determinant may be related to 
determinants of {\it finite} size matrices~: all above expressions have
involved {\it arc} matrices of size $2B\times2B$. 
We will now see that we can express the spectral
determinant in terms of $V\times V$ vertex matrices in general
smaller\footnote{The spectral determinant can only be expressed in
  term of vertex matrices when boundary conditions at the vertices are
  such that it is possible to assign a unique variable to each
  vertex. This can only be done for ``permutation invariant'' boundary
  conditions~\cite{Des01}. Two such particular cases are discussed in
  the two following sections.}.

\section{Continuous boundary conditions}
\label{sec:cbc}

\subsection{Roth's trace formula}

Let us consider the important case of the Laplace operator acting on
functions that are continuous at the vertices~:
$\psi_\ab(0)=\psi_\alpha$ for all vertices $\beta$ neighbours of
$\alpha$, and 
$\sum_\beta{}a_\ab{}(\Dc_x\psi)_\ab(0)=\lambda_\alpha\psi_\alpha$, where
the connectivity matrix ensures that the sum runs over vertices
neighbours of $\alpha$.
The parameter $\lambda_\alpha$ must be chosen real in order to ensure
self adjointness of the Schr\"odinger operator~; it can be
understood as the weight of a $\delta$ potential at the vertex (these
boundary conditions are sometimes denoted as
``$\delta$-coupling''~\cite{Exn95,Exn97b}). 
In the appropriate basis where arcs are gathered by vertices from
which they issue, the matrices $C$ and $D$ have block diagonal
structures (with $V$ blocks $\alpha=1,\cdots{}V$) where the
$m_\alpha\times{}m_\alpha$ block $\alpha$ correspond to the vertex. 
A set of $\gamma$-independent matrices $C$ and $D$ corresponding to
$\delta$-couplings are made of $m_\alpha\times{}m_\alpha$ blocks 
\begin{equation}
  \label{eq:CDcont}
  C^{(\alpha)}  =  
  \left(
    \begin{array}{ccccc}
      -\lambda_\alpha & 0 & 0 & \cdots & 0\\
      -1 & 1 & 0 & \cdots & 0\\
      -1 & 0 & 1 & \cdots & 0\\
      \vdots & \vdots &  & & \vdots \\
      -1 & 0 & 0 & \cdots  & 1
    \end{array}
  \right)
  \hspace{0.5cm} \mbox{and} \hspace{0.5cm}
  D^{(\alpha)} =
  \left(
    \begin{array}{ccccc}
     1 & 1 & \cdots & 1 & 1\\
     0 & 0 & \cdots & 0 & 0 \\
     0 & 0 & \cdots & \cdots & 0 \\
     \vdots & \vdots &  & & \vdots \\
     0 & 0 & \cdots & \cdots & 0
    \end{array}
  \right)
\end{equation}
It will be useful to remark that
$\det(C^{(\alpha)}-\sqrt\gamma\,D^{(\alpha)})=-(\lambda_\alpha+m_\alpha\sqrt\gamma)$.  
The vertex ``scattering matrix'' for the vertex
$Q^{(\alpha)}=(-C^{(\alpha)}+\sqrt\gamma\,D^{(\alpha)})^{-1}(C^{(\alpha)}+\sqrt\gamma\,D^{(\alpha)})$
takes the form~\cite{AkkComDesMonTex00} 
\begin{equation}
  \label{eq:QBC}
  Q^{(\alpha)} = \frac{2}{m_\alpha+\lambda_\alpha/\sqrt\gamma}
  \begin{pmatrix}
    1&1&\cdots&1\\
    1&1&\cdots&1\\
    \vdots&\vdots& &\vdots\\
    1&1&\cdots&1
  \end{pmatrix}
  -\mathbf{1}_{m_\alpha}
  \:,
\end{equation}
(for $\gamma=0$ we recognise the weights introduced by
Roth~\cite{Rot83}~:
$\frac2{m_\alpha}$ for visiting the vertex $\alpha$ and
$\frac2{m_\alpha}-1$ for a reflection on it).
Note that the limit $\lambda_\alpha=\infty$ corresponds to Dirichlet boundary
conditions $\psi_i(0)=0$.

Moreover let us consider the case $V(x)=0$, then 
$r_a=0$ and $t_a=\EXP{-\sqrt\gamma\,l_a}$
(note that this latter is the analytic continuation of the
transmission probability amplitude through a free interval
$t_a=\EXP{\I{}kl_a}$ for $\sqrt{\gamma}=\sqrt{-k^2-\I0^+}$).
Starting from \eqref{eq:Sgbc2JD} and using
\eqref{eq:useful6} we get
\begin{equation}
  \label{eq:LfctCB}
  S(\gamma) = 
  2^{-B}\prod_\alpha\left(m_\alpha+\frac{\lambda_\alpha}{\sqrt\gamma}\right)\:
  \gamma^{\frac{V-B}{2}}\EXP{\sqrt\gamma\mathcal{L}} 
  \underbrace{
    \prod_{\tc}
    \left( 1 - \alpha(\tc)\,
      \EXP{ - \sqrt{\gamma}\ell(\tc)
            + \I\theta(\tc) } 
    \right)
  }_{\det(\mathbf{1}_{2B}-QR)}
\end{equation}
where the first product runs over the $V$ vertices.
$\mathcal{L}=\sum_{(a)}l_a$ is the total length of the graph.
The weights of the orbit $\mathcal{C}=(a_1,a_2,\cdots,a_n)$ is now denoted
$v(\mathcal{C})\to\alpha(\mathcal{C})=Q_{a_1\bar{a}_2}Q_{a_2\bar{a}_3}\cdots{}Q_{a_n\bar{a}_1}$
for $Q$ given by gathering the blocks \eqref{eq:QBC}.
The notations 
$\ell(\mathcal{C})$ and $\theta(\mathcal{C})$ designate  the length of
the orbit and the magnetic flux enclosed by it, respectively.
Thanks to the simple structure of the matrix $R$ for $V(x)=0$, the
bond scattering part of the weight of the orbit is simply 
$b(\mathcal{C})\to\EXP{-\sqrt{\gamma}\ell(\mathcal{C})+\I\theta(\mathcal{C})}$.

If we choose moreover vertex scattering with $\lambda_\alpha=0$, the
weights $\alpha(\tc)$ become $\gamma$-independent, what allows a
simple Laplace inversion of the derivative
$\derivp{}{\gamma}\ln{}S(\gamma)=\int_0^\infty\D{}t\,Z(t)\,\EXP{-\gamma{}t}$. We
recover the Roth's trace formula~\cite{Rot83}~:
\begin{equation}
  \label{Roth}
  \boxed{
  Z(t)=
  \frac{\mathcal{L}}{2\sqrt{\pi t}} + \frac{V-B}{2} 
  + \frac{1}{2\sqrt{\pi t}}
  \sum_{\mathcal{C}} \ell(\tc) \alpha(\mathcal{C})
  \EXP{-\frac{\ell(\mathcal{C})^2}{4t}+\I\theta(\mathcal{C})}  
  }
  \hspace{0.5cm}\mbox{for } \lambda_\alpha=0\:\forall\:\alpha
  \:,
\end{equation} 
where the sum now runs over all periodic orbits $\mathcal{C}$,
where $\tc$ is the primitive orbit related to the orbit~$\mathcal{C}$
(see Refs.~\cite{Rot83,AkkComDesMonTex00} for more details).

\subsection{Vertex determinant}

Another interesting expression of the
spectral determinant in terms of $V\times{}V$-matrix coupling
vertices may be obtained by a construction anologous to the one given
above.
The asumption that the functions are continuous at the vertices allows
us to deal with vertex variables (see figure~\ref{fig:vertex}.b)~:
\begin{align}
  \label{eq:8}
  \mathcal{G}_{\alpha\beta,\mu\nu}(x_{\alpha\beta},y_{\mu\nu})
&= 
\mathcal{G}_\alpha\,f_{\alpha\beta}(x_{\alpha\beta})\,
\EXP{\I A_{\alpha\beta}x_{\alpha\beta}}
+ \mathcal{G}_\beta\,f_{\beta\alpha}(x_{\beta\alpha})\,
\EXP{\I A_{\beta\alpha}x_{\beta\alpha}}
\nonumber\\
&+ \delta_{(\mu\nu),(\alpha\beta)}
 \frac{\EXP{\I A_{\alpha\beta}(x_{\alpha\beta}-y_{\alpha\beta})} }
        {{W}_{\alpha\beta}}\:
f_{\alpha\beta}(\max{x_{\alpha\beta}}{y_{\alpha\beta}})\:
f_{\beta\alpha}(\max{x_{\beta\alpha}}{y_{\beta\alpha}})
\end{align}
where $\mathcal{G}_\alpha$ is the value of the Green function at the vertex.
The boundary condition at vertex $\mu$ takes the form
\begin{equation}
\sum_{\nu}\mathcal{M}_{\mu\nu}\,\mathcal{G}_\nu = 
\delta_{\mu\alpha}\,f_{\alpha\beta}(y_{\alpha\beta})\,
\EXP{-\I A_{\alpha\beta}y_{\alpha\beta}}
+\delta_{\mu\beta}\,f_{\beta\alpha}(y_{\beta\alpha})\,
\EXP{-\I A_{\beta\alpha}y_{\beta\alpha}}
\end{equation}
where 
the $V\times V$
matrices coupling vertices is given by~:
\begin{equation}
  \mathcal{M}_\ab = 
  \delta_\ab
  \left(\lambda_\alpha - \sum_\mu a_{\alpha\mu}f'_{\alpha\mu}(0)\right)
  +a_\ab\,f'_\ba(l_\ab)\,\EXP{-\I\theta_\ab}
  \label{eq:Jean1}
  \:.
\end{equation}
We can now obtain the value of the Green function at the vertices and
reinject these expressions in \eqref{eq:8}. Integration of 
\begin{align}
\mathcal{G}_{\alpha\beta,\alpha\beta}(x_{\alpha\beta},x_{\alpha\beta}) 
 &= 
  \left(\mathcal{M}^{-1}\right)_{\alpha\alpha}\,
  f_{\alpha\beta}(x_{\alpha\beta})^2
+ \left[
      \EXP{\I\theta_\ab}\left(\mathcal{M}^{-1}\right)_{\alpha\beta}
    + \EXP{-\I\theta_\ab}\left(\mathcal{M}^{-1}\right)_{\beta\alpha}
  \right]
  f_{\alpha\beta}(x_{\alpha\beta})\,f_{\beta\alpha}(x_{\beta\alpha})
\nonumber\\
&+ \left(\mathcal{M}^{-1}\right)_{\beta\beta}\,
  f_{\beta\alpha}(x_{\beta\alpha})^2
+ \frac{1}{{W}_\ab}
  f_{\alpha\beta}(x_{\alpha\beta})\,f_{\beta\alpha}(x_{\beta\alpha})
\end{align}
can be performed along the same lines as before. One obtains~\cite{Des00}
\begin{equation}
  \label{eq:Jean00}
  \boxed{
  S(\gamma)=\prod_{(\ab)}[-f_\ab'(l_\ab)]^{-1}\:\det\mathcal{M}
  }
\end{equation}
where the product runs over the bonds of the graph.
Eq.~\eqref{eq:Jean00} is less general than Eq.~\eqref{eq:Sgbc1JD}
since it corresponds to a particular choice of boundary conditions at
the vertices,
however it involves the information about the graph in a more compact
manner in the sense that the $V\times V$ matrix $\mathcal{M}$ is
smaller than the $2B\times2B$ matrix $C+DM$. 

In the absence of a potential, $V(x)=0$, we recover
Pascaud \& Montambaux' result~\cite{PasMon99}
\begin{equation}
  \label{eq:spedet}
  S(\gamma) = 
  \prod_{(\ab)}\frac{\sinh\sqrt\gamma l_\ab}{\sqrt\gamma}\:
  \det\mathcal{M}
  \:,
\end{equation}
for 
\begin{equation}
  \mathcal{M}_\ab = 
  \delta_\ab
  \left(\lambda_\alpha +\sqrt{\gamma} \sum_\mu a_{\alpha\mu}
  \coth\sqrt{\gamma}l_{\alpha\mu}\right)
  -a_\ab
  \frac{\sqrt\gamma\:\EXP{-\I\theta_\ab}}{\sinh\sqrt{\gamma}l_\ab}
  \:.\label{eq:PasMon}
\end{equation}
Note the alternative derivation of \eqref{eq:spedet} and
\eqref{eq:Jean00} with the path integral~\cite{AkkComDesMonTex00} and
\cite{Des00a}.  
It is worth pointing that the full spectrum of the Laplace operator is
not always given by $\det\mathcal{M}=0$~: it may occur in some
particular cases
that eigenstates of the Schr\"odinger operator vanishes at all
vertices $\psi_\alpha=0$ for $\psi(x)\neq0$. Some examples are discussed
in details in Ref.~\cite{Tex02}
(this is related to the phenomenon known as ``bound state in the
continuum'' in the scattering situation, for graphs with some
infinitly long wires \cite{NeuWig29,StiHer75}). 

Finally, note that 
the $\gamma$-independent prefactor in \eqref{eq:Jean00} is chosen in
order to match with formulae \eqref{eq:Sgbc1JD} and
\eqref{eq:Sgbc2JD}~: one can easily check that the $\gamma\to\infty$ 
behaviours coincide. 
Note that, in Ref.~\cite{AkkComDesMonTex00}, a direct relation between
\eqref{eq:spedet} and \eqref{eq:LfctCB} was established, without {\it
  a posteriori} matching procedure.

\section{Continuous derivative boundary conditions}
\label{sec:cdbc}

\subsection{Trace formula}

Another interesting case worth to be discussed is the case of
Schr\"odinger operator acting on functions
whose derivative are continuous at the vertices
$(\Dc_x\psi)_\ab(0)=\psi'_\alpha$ for all vertices $\beta$ neighbours of
$\alpha$, and 
$\sum_\beta{}a_\ab\psi_\ab(0)=\mu_\alpha\psi'_\alpha$.
This choice is denoted as ``$\delta'_s$-coupling'' in
Refs.~\cite{Exn95,Exn97b}.
A set of $\gamma$-independent boundary matrices at vertex $\alpha$ are
in this case 
\begin{equation}
  \label{eq:CDcontder}
  C^{(\alpha)}  =  
  \left(
    \begin{array}{ccccc}
     1 & 1 & \cdots & 1 & 1\\
     0 & 0 & \cdots & 0 & 0 \\
     0 & 0 & \cdots & \cdots & 0 \\
     \vdots & \vdots &  & & \vdots \\
     0 & 0 & \cdots & \cdots & 0
    \end{array}
  \right)
  \hspace{0.5cm} \mbox{and} \hspace{0.5cm}
  D^{(\alpha)} =
  \left(
    \begin{array}{ccccc}
      -\mu_\alpha & 0 & 0 & \cdots & 0\\
      -1 & 1 & 0 & \cdots & 0\\
      -1 & 0 & 1 & \cdots & 0\\
      \vdots & \vdots &  & & \vdots \\
      -1 & 0 & 0 & \cdots & 1
    \end{array}
  \right)
\end{equation}
that gives 
$\det(C^{(\alpha)}-\sqrt\gamma\,D^{(\alpha)})=(-1)^{m_\alpha+1}\gamma^{m_\alpha/2}(\mu_\alpha+m_\alpha/\sqrt\gamma)$.  
From \eqref{eq:Qmatrix}, we get the vertex ``scattering matrix'' for the
vertex $\alpha$~:
\begin{equation}
  \label{eq:Qcd}
  Q^{(\alpha)} = \frac{-2}{m_\alpha+\mu_\alpha\sqrt\gamma}
  \begin{pmatrix}
    1&1&\cdots&1\\
    1&1&\cdots&1\\
    \vdots&\vdots& &\vdots\\
    1&1&\cdots&1
  \end{pmatrix}
  +\mathbf{1}_{m_\alpha}
\end{equation}
therefore the limit $\mu_\alpha=\infty$ corresponds to Neumann boundary
conditions $\psi_i'(0)=0$.

Starting again from Eq.~\eqref{eq:Sgbc2JD}, a little bit of algebra gives the
spectral determinant in the absence of potential
\begin{equation}
  \label{eq:TraceScd}
  S(\gamma) = 
  2^{-B}\prod_\alpha\left(m_\alpha+\sqrt\gamma\mu_\alpha\right)\:
  \gamma^{\frac{B-V}{2}}\EXP{\sqrt\gamma\mathcal{L}} 
    \prod_{\tc}
    \left( 1 - \beta(\tc)\,
      \EXP{ - \sqrt{\gamma}\ell(\tc)
            + \I\theta(\tc) } 
    \right)
\end{equation}
where the weights $v(\mathcal{C})\to\beta(\mathcal{C})$ are computed
with matrix elements of \eqref{eq:Qcd}.
It is interesting to point that these weights may be simply related to
the weights $\alpha(\mathcal{C})$ of the 
previous section thanks to the substitution
$\lambda_\alpha\to\gamma\mu_\alpha$ and adding 
a sign $(-1)^{\#\,\mathrm{arcs\:of\:}\mathcal{C}}$.

If we moreover set $\mu_\alpha=0$ we may obtain the trace formula for
the partition function
\begin{equation}
  \label{eq:TraceZcd}
  \boxed{
  Z(t)=
  \frac{\mathcal{L}}{2\sqrt{\pi t}} + \frac{B-V}{2} 
  + \frac{1}{2\sqrt{\pi t}}
  \sum_{\mathcal{C}} \ell(\tc) \beta(\mathcal{C})
  \EXP{-\frac{\ell(\mathcal{C})^2}{4t}+\I\theta(\mathcal{C})}  
  }
  \hspace{0.5cm}\mbox{for } \mu_\alpha=0\:\forall\:\alpha
  \:,
\end{equation} 
that generalises the result of Roth to the case of the Laplace
operator acting on functions whose derivative is continuous at the vertices.
Compare to the case of functions continuous at the vertices discussed
in the previous section, Eq.~\eqref{Roth}, only the constant term and
the weights of the orbits have changed in sign.

\subsection{Vertex determinant}

The expression of the spectral determinant may also be obtained 
by using vertex variables~\cite{Tex08}~:
\begin{equation}
  \label{eq:spedetcd}
  \boxed{
  S(\gamma)=\prod_{(\ab)}[-g_\ab(l_\ab)]^{-1}\det\mathcal{N}
  }
\end{equation}
with
\begin{equation}
  \label{eq:2}
  \mathcal{N}_\ab = 
  \delta_\ab\left(\mu_\alpha - \sum_\nu a_{\alpha\nu}g_{\alpha\nu}(0)\right)
  -a_\ab\, g_\ab(l_\ab)\,\EXP{-\I\theta_\ab}
\end{equation}
where the functions $g_a$ were introduced above (and in the  appendix~\ref{sec:bondscatt}).
In the absence of the potential, $V(x)=0$, we obtain
\begin{equation}
  S(\gamma)=\prod_{(\ab)}\sqrt\gamma\sinh\sqrt{\gamma}l_\ab\:\det\mathcal{N}
\end{equation}
where
$
  \mathcal{N}_\ab = 
  \delta_\ab
  \left(\mu_\alpha +\frac1{\sqrt{\gamma}} \sum_\nu a_{\alpha\nu}
  \coth\sqrt{\gamma}l_{\alpha\nu}\right)
  +a_\ab \frac{\EXP{-\I\theta_\ab}}{\sqrt\gamma\sinh\sqrt{\gamma}l_\ab}
$.

\section{From the trace formula to the $\zeta$-regularised determinant}
\label{sec:zr}

The question of the prefactor of the spectral determinant 
is the subject of the present section. 
It is important to fix the $\gamma$-independent prefactor~:
in all expressions given in the previous sections, 
\eqref{eq:Sgbc1JD},
\eqref{eq:DualJean},  \eqref{eq:Sgbc2JD}, \eqref{eq:LfctCB},  \eqref{eq:spedet},
\eqref{eq:TraceScd} and  \eqref{eq:spedetcd}, 
the prefactor has been chosen arbitrarily, nevertheless in a way that all
expressions match when varying continuously boundary
conditions. However this choice is {\it a priori} not related to a
particular regularisation and has only been made for convenience.

\subsection{Functions continuous at the vertices}

Let us first consider the simplest case of the Laplace operator 
acting on functions continuous at the vertices in the absence of
potential. Moreover we set the parameters $\lambda_\alpha=0$, when the
Roth's trace formula for the partition function holds.
Mellin transform of the Roth's partition function \eqref{Roth} reads
\begin{align}
  \zeta(s,\gamma) &=
  -\mathcal{L}\,\frac{\Gamma(s-1/2)}{\Gamma(-1/2)\Gamma(s)}\,\gamma^{-s+1/2} 
  + \frac{V-B}{2}\, \gamma^{-s} 
  \nonumber\\
  &\hspace{1cm}
  + \frac{1}{\sqrt{\pi}\Gamma(s)}\, \gamma^{-s+1/2} 
  \sum_{\mathcal{C}} \ell(\tc) \alpha(\mathcal{C})
  \EXP{\I\theta(\mathcal{C})}  
  \left(\frac2{\sqrt\gamma\,\ell(\mathcal{C})}\right)^{-s+1/2}\,
  K_{\frac12-s}\left(\sqrt\gamma\,\ell(\mathcal{C})\right)
\end{align}
where $K_\nu(z)$ is the MacDonald function (modified Bessel
function)~\cite{gragra}.
First and third terms are of the form
$\frac1{\Gamma(s)}h(s)$ for a function $h(s)$ regular for
$s\to0$, therefore we can use $\big[\frac1{\Gamma(s)}h(s)\big]'_0=h(0)$.
The analysis of the third term requires the formula
$\big[\derivp{K_\nu(z)}{\nu}\big]_{\nu=1/2}=-\sqrt{\frac{\pi}{2z}}\,\EXP{z}\,\mathrm{Ei}(-2z)$~\cite{gragra}.
Using  and 
$K_{1/2}(z)=\sqrt{\frac{\pi}{2z}}\,\EXP{-z}$ we get
\begin{equation}
  \deriv{\zeta}{s}(0,\gamma) = 
  -\sqrt\gamma\, \mathcal{L} -\frac{V-B}{2}\, \ln\gamma
  +\sum_{\mathcal{C}} 
   \frac{\ell(\tc) \alpha(\mathcal{C})}{\ell(\mathcal{C})}\,
  \EXP{-\sqrt\gamma\,\ell(\mathcal{C})+\I\theta(\mathcal{C})}
\end{equation}
Sum over orbits can be decomposed as a sum over primitive orbits and
their repetitions $\sum_{\mathcal{C}}=\sum_{\tc}\sum_{n=1}^\infty$.
Using $\ell(\mathcal{C})=n\,\ell(\tc)$,
$\theta(\mathcal{C})=n\,\theta(\tc)$ and
$\alpha(\mathcal{C})=\alpha(\tc)^n$.
Therefore
\begin{equation}
  \deriv{\zeta}{s}(0,\gamma) = 
  -\sqrt\gamma\, \mathcal{L} -\frac{V-B}{2}\, \ln\gamma
  -\sum_{\tc}
  \ln\left( 
     1 - \alpha(\tc)\,
     \EXP{-\sqrt\gamma\, \ell(\tc)+\I\theta(\tc)} 
  \right)
\end{equation}
that is
\begin{equation}
  \boxed{
     S^\zeta(\gamma) = \gamma^{\frac{V-B}2}\EXP{\sqrt\gamma\,\mathcal{L}}
    \prod_{\tc}
    \left( 1 - \alpha(\tc)\,
      \EXP{ - \sqrt{\gamma}\ell(\tc)
            + \I\theta(\tc) } 
    \right)
  = \gamma^{\frac{V-B}2}\EXP{\sqrt\gamma\,\mathcal{L}}\,\det(\mathbf{1}_{2B}-QR)
  }
\end{equation}
We recognise part of the expression \eqref{eq:LfctCB} (see also
Eq.~(65) of Ref.~\cite{AkkComDesMonTex00}). 
We have related above Eq.~\eqref{eq:LfctCB} to another representation in
terms of a vertex matrix, Eq.~\eqref{eq:spedet}, hence
\begin{equation}
  \label{eq:SzetaReg}
   S^\zeta(\gamma) = 
   \prod_{(\ab)}\frac{2\sinh\sqrt\gamma l_\ab}{\sqrt\gamma}\:
   \frac{ \det\mathcal{M} }{ \prod_\alpha m_\alpha }
  =\frac{2^B}{(\prod_\alpha m_\alpha)}S(\gamma) 
  \hspace{0.5cm}\mbox{for }
  V(x)=0 \ \& \ \lambda_\alpha=0\ \forall\,\alpha
  \:.
\end{equation}

The previous formula may be easily generalised to the case of the
Schr\"odinger equation for continuous boundary condition with
arbitrary parameters $\lambda_\alpha$.
Using the observation that various regularisations only differ
by a $\gamma$-independent prefactor, we conclude that
$S^\zeta(\gamma)=2^B/(\prod_\alpha m_\alpha)\,S(\gamma)$
also holds in the presence of the potential
\begin{equation}
  \label{eq:ZetaSCB}
  \boxed{
    S^\zeta(\gamma) =
    \prod_{(\ab)}\frac{-2}{f_\ab'(l_\ab)}\:
    \frac{\det\mathcal{M}}{\prod_\alpha{}m_\alpha}
  }
\end{equation}
This is precisely Friedlander's result \cite{Fri06} 
({\it cf.} Ref.~\cite{Tex08}, where the correspondence
of notations is discussed).
The product $\prod_{(\ab)}[{-2}/{f_\ab'(l_\ab)}]$ is the Dirichlet
determinant of the graph (determinant when Dirichlet boundary
conditions are chosen at all vertices that make all wires independent).

\paragraph{Example 1 : wire}

Let us consider a wire of length $L$ ($B=1$ bond and $V=2$ vertices)
with no potential $V(x)=0$.
Matrices $C$ and $D$ for boundary conditions of type~\eqref{eq:CDcont}
are
\begin{equation}
  C =
  \begin{pmatrix}
    -\lambda_1 & 0 \\
    0 & -\lambda_2
  \end{pmatrix}
  \hspace{0.25cm}\mbox{and}\hspace{0.25cm}
  D = \begin{pmatrix}
    1 & 0 \\ 0 & 1
  \end{pmatrix}
\end{equation}
where $\lambda_1$ and $\lambda_2$ describe the boundary conditions at
the two vertices.
Eq.~\eqref{eq:conjecture} gives
\begin{equation}
  S^\zeta_\mathrm{wire}(\gamma) = 
  2\sqrt\gamma\,\sinh\sqrt\gamma L 
  +(\lambda_1+\lambda_2)2\cosh\sqrt\gamma L
  +\lambda_1\lambda_2\frac{2\sinh\sqrt\gamma L}{\sqrt\gamma}
\end{equation}
This result can be obtained straightforwardly from \eqref{eq:ZetaSCB}.

For $\lambda_1=\lambda_2=0$ we recover the Neumann determinant
$S^\zeta_\mathrm{wire}(\gamma)=2\sqrt\gamma\,\sinh\sqrt\gamma
L=2\gamma L\prod_{n=1}^\infty\big(1+\frac{\gamma}{E_n}\big)$
where $E_n=\big(\frac{n\pi}{L}\big)^2$ with $n\in\mathbb{N}$ are the
eigenvalues of the Laplace operator on $(0,L)$ for Neumann boundary
conditions. 

Note that the second term $2\cosh\sqrt\gamma L$ corresponds to the
Neumann/Dirichlet determinant (Neumann at $x=0$ and Dirichlet at $x=L$
or vice versa) retained for $\lambda_1=0$ and $\lambda_2=\infty$ (with
$\lambda_1\lambda_2=0$).

The last term $\frac{2\sinh\sqrt\gamma L}{\sqrt\gamma}$ is the
Dirichlet determinant ($\lambda_1=\lambda_2=\infty$).

\paragraph{Example 2 : ring}

We consider a ring  of perimeter $L$ ($B=1$ bond and $V=1$ vertex)
with no potential $V(x)=0$ and pierced by a flux $\theta$.
Matrices $C$ and $D$ for boundary conditions of type~\eqref{eq:CDcont}
are
\begin{equation}
  C =
  \begin{pmatrix}
    -\lambda & 0 \\
    -1 & 1
  \end{pmatrix}
  \hspace{0.25cm}\mbox{and}\hspace{0.25cm}
  D = \begin{pmatrix}
    1 & 1 \\ 0 & 0
  \end{pmatrix}
\end{equation}
therefore \eqref{eq:conjecture} gives
\begin{equation}
  S^\zeta_\mathrm{ring}(\gamma) = 
  2\, \big( \cosh\sqrt\gamma L - \cos \theta\big)
  +\lambda\frac{\sinh\sqrt\gamma L}{\sqrt\gamma}
\end{equation}
It may also be obtained from \eqref{eq:ZetaSCB} knowing how a loop
contributes to $\mathcal{M}$ (see Ref.~\cite{AkkComDesMonTex00}).

For $\lambda=0$ we obtain
$S^\zeta_\mathrm{ring}(\gamma)=2(\cosh\sqrt\gamma L-\cos\theta)=4\sin^2\frac\theta2\prod_{n\in\mathbb{Z}}\big(1+\frac{\gamma}{E_n}\big)$
where 
 $E_n=\big(\frac{2n\pi-\theta}{L}\big)^2$ with $n\in\mathbb{Z}$ is the
 spectrum of the ring.

\paragraph{Example 3 : star graph}

We consider a star graph with $B$ arms (then $V=B+1$)
with no potential $V(x)=0$.
We introduce a parameter $\lambda$ at the central vertex and set all
other boundary parameters equal to zero, $\lambda_\alpha=0$.
In this case it is more easy to use \eqref{eq:SzetaReg}. The $B$ wires
are ``dead arms'' and may be taken into account very simply through the
following rule \cite{Pas98}~:
a dead arm issuing from the vertex $\alpha$ gives a contribution
$\sqrt\gamma\tanh\sqrt\gamma l_a$ that should be added to the matrix
element $\mathcal{M}_{\alpha\alpha}$ corresponding to the graph from
which the dead 
arm is removed~; in the product over bonds of Eq.~\eqref{eq:SzetaReg}
we replace $\frac{\sinh\sqrt\gamma l_a}{\sqrt\gamma}$ by 
$\cosh\sqrt\gamma l_a$.
This simple rule can be easily recovered using the path integral
formalism introduced in Ref.~\cite{AkkComDesMonTex00}.
It allows to treat the star graph as a graph with one vertex
and $B$ dead arms. Therefore
$\mathcal{M}=\lambda+\sqrt\gamma\sum_{a=1}^B\tanh\sqrt\gamma l_a$ and
the spectral determinant is given by 
\begin{equation}
   S^\zeta_\mathrm{star}(\gamma) = \frac{2^B}{B}
  \prod_{a=1}^B\cosh\sqrt\gamma l_a
  \left(
    \lambda + \sqrt\gamma\sum_{a=1}^B\tanh\sqrt\gamma l_a
  \right)
  \:.
\end{equation}
If we consider the case $\lambda=0$ and take the limit $\gamma\to0$ we
recover the result of Ref.~\cite{HarKir09}
\begin{equation}
    S^\zeta_\mathrm{star}(\gamma) \underset{\gamma\to0}{\simeq} 
    \frac{2^B\mathcal{L}}{B}\gamma = \gamma\: {\det}'(-\Delta)
\end{equation}
where $\det'(-\Delta)$ is the determinant when the zero mode
contribution has been excluded.

\subsection{Functions with derivative continuous at the vertices}

The reader has remarked that there is very little difference between the
Roth trace formula \eqref{Roth} and its generalisation \eqref{eq:TraceZcd}~: 
the second term proportional to $V-B$ is changed in sign and the
definition of the weights are changed~:
when $\lambda_\alpha=0$ and $\mu_\alpha=0$ we have simply
$\beta(\tc)=(-1)^{\#\,\mathrm{arcs\:of\:}\tc}\alpha(\tc)$.
We can therefore perform a similar calculation and get in the case 
$V(x)=0$ and $\mu_\alpha=0$~:
\begin{equation}
     S^\zeta(\gamma) = \gamma^{\frac{B-V}2}\EXP{\sqrt\gamma\,\mathcal{L}}
    \prod_{\tc}
    \left( 1 - \beta(\tc)\,
      \EXP{ - \sqrt{\gamma}\ell(\tc)
            + \I\theta(\tc) } 
    \right)
\end{equation}
By a similar continuity argument as in the previous subsection we get
for $V(x)\neq0$ and $\mu_\alpha\neq0$~: 
\begin{equation}
  \label{eq:ZetaSCDB}
  \boxed{
    S^\zeta(\gamma) =
    \prod_{(\ab)}\frac{-2}{g_\ab(l_\ab)}\:
    \frac{\det\mathcal{N}}{\prod_\alpha{}m_\alpha}
  }
\end{equation}
the first term (product over the bonds) coincides now with the Neumann
determinant of the graph (product of spectral deterimants for all
disconnected wires with Neumann boundary conditions).
This latter expression hence provides some generalisation of
Friedlander's result to another kind of boundary conditions.


\section{Conclusion}

Using the Roth's trace formula for the partition function and its
generalisation, we have obtained the expression of the
$\zeta$-regularised spectral determinant of the Schr\"odinger operator
acting on functions continuous at the vertices,
Eq.~\eqref{eq:ZetaSCB}, or functions whose derivative is continuous,
Eq.~\eqref{eq:ZetaSCDB}. 
It is worth emphasizing that in these formulae, the
$\gamma$-independent prefactor is now fully determined by some
properties of the Graph~: i.e. the $\zeta$-regularisation {\it fixes a 
  priori} the prefactor of the spectral determinant, whereas in
formulae \eqref{eq:Sgbc1JD}, \eqref{eq:DualJean},  \eqref{eq:Sgbc2JD},
\eqref{eq:LfctCB},  \eqref{eq:spedet}, \eqref{eq:TraceScd} and
\eqref{eq:spedetcd} it was {\it chosen} arbitrarily {\it a posteriori}.

\begin{figure}[!ht]
  \centering
\begin{picture}(0,0)%
\epsfig{file=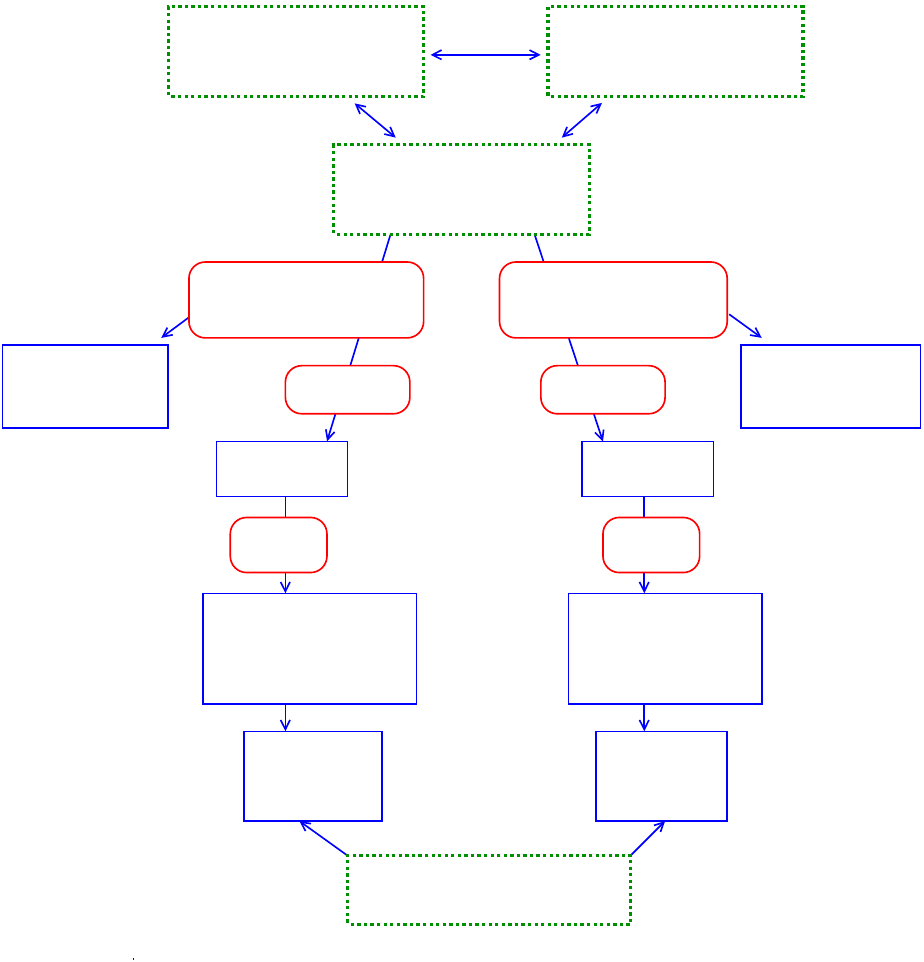}%
\end{picture}%
\setlength{\unitlength}{4144sp}%
\begingroup\makeatletter\ifx\SetFigFont\undefined%
\gdef\SetFigFont#1#2#3#4#5{%
  \reset@font\fontsize{#1}{#2pt}%
  \fontfamily{#3}\fontseries{#4}\fontshape{#5}%
  \selectfont}%
\fi\endgroup%
\begin{picture}(4221,4388)(2541,-4639)
\put(3531,-3081){\makebox(0,0)[lb]{\smash{{\SetFigFont{8}{9.6}{\rmdefault}{\mddefault}{\updefault}{\color[rgb]{0,0,0}Trace formula}%
}}}}
\put(3531,-3239){\makebox(0,0)[lb]{\smash{{\SetFigFont{8}{9.6}{\rmdefault}{\mddefault}{\updefault}{\color[rgb]{0,0,0}for $Z(t)$ (Roth)}%
}}}}
\put(3531,-3397){\makebox(0,0)[lb]{\smash{{\SetFigFont{8}{9.6}{\rmdefault}{\mddefault}{\updefault}{\color[rgb]{0,0,0}Eq. \eqref{Roth}}%
}}}}
\put(3720,-3902){\makebox(0,0)[lb]{\smash{{\SetFigFont{8}{9.6}{\rmdefault}{\mddefault}{\updefault}{\color[rgb]{0,0,0}Eq. \eqref{eq:ZetaSCB}}%
}}}}
\put(3783,-3744){\makebox(0,0)[lb]{\smash{{\SetFigFont{8}{9.6}{\rmdefault}{\mddefault}{\updefault}{\color[rgb]{0,0,0}$S^\zeta$}%
}}}}
\put(5330,-3902){\makebox(0,0)[lb]{\smash{{\SetFigFont{8}{9.6}{\rmdefault}{\mddefault}{\updefault}{\color[rgb]{0,0,0}Eq. \eqref{eq:ZetaSCDB}}%
}}}}
\put(5393,-3744){\makebox(0,0)[lb]{\smash{{\SetFigFont{8}{9.6}{\rmdefault}{\mddefault}{\updefault}{\color[rgb]{0,0,0}$S^\zeta$}%
}}}}
\put(3373,-609){\makebox(0,0)[lb]{\smash{{\SetFigFont{8}{9.6}{\rmdefault}{\mddefault}{\updefault}{\color[rgb]{0,0,0}$S\propto\det(C+DM)$}%
}}}}
\put(3373,-430){\makebox(0,0)[lb]{\smash{{\SetFigFont{8}{9.6}{\rmdefault}{\mddefault}{\updefault}{\color[rgb]{0,0,0}Eq. \eqref{eq:Sgbc1JD}}%
}}}}
\put(5109,-620){\makebox(0,0)[lb]{\smash{{\SetFigFont{8}{9.6}{\rmdefault}{\mddefault}{\updefault}{\color[rgb]{0,0,0}$S\propto\det(CN+D)$}%
}}}}
\put(5109,-430){\makebox(0,0)[lb]{\smash{{\SetFigFont{8}{9.6}{\rmdefault}{\mddefault}{\updefault}{\color[rgb]{0,0,0}Eq. \eqref{eq:DualJean}}%
}}}}
\put(4162,-1061){\makebox(0,0)[lb]{\smash{{\SetFigFont{8}{9.6}{\rmdefault}{\mddefault}{\updefault}{\color[rgb]{0,0,0}Eq. \eqref{eq:Sgbc2JD}}%
}}}}
\put(4162,-1251){\makebox(0,0)[lb]{\smash{{\SetFigFont{8}{9.6}{\rmdefault}{\mddefault}{\updefault}{\color[rgb]{0,0,0}$S\propto\det(1-QR)$}%
}}}}
\put(4162,-4343){\makebox(0,0)[lb]{\smash{{\SetFigFont{8}{9.6}{\rmdefault}{\mddefault}{\updefault}{\color[rgb]{0,0,0}Conjecture : Eq. \eqref{eq:conjecture}}%
}}}}
\put(5172,-3081){\makebox(0,0)[lb]{\smash{{\SetFigFont{8}{9.6}{\rmdefault}{\mddefault}{\updefault}{\color[rgb]{0,0,0}Trace formula}%
}}}}
\put(5172,-3239){\makebox(0,0)[lb]{\smash{{\SetFigFont{8}{9.6}{\rmdefault}{\mddefault}{\updefault}{\color[rgb]{0,0,0}for $Z(t)$}%
}}}}
\put(5172,-3397){\makebox(0,0)[lb]{\smash{{\SetFigFont{8}{9.6}{\rmdefault}{\mddefault}{\updefault}{\color[rgb]{0,0,0}Eq. \eqref{eq:TraceZcd}}%
}}}}
\put(5992,-1945){\makebox(0,0)[lb]{\smash{{\SetFigFont{8}{9.6}{\rmdefault}{\mddefault}{\updefault}{\color[rgb]{0,0,0}Eq. \eqref{eq:spedetcd}}%
}}}}
\put(5992,-2134){\makebox(0,0)[lb]{\smash{{\SetFigFont{8}{9.6}{\rmdefault}{\mddefault}{\updefault}{\color[rgb]{0,0,0}$S\propto\det\mathcal{N}$}%
}}}}
\put(3468,-1598){\makebox(0,0)[lb]{\smash{{\SetFigFont{8}{9.6}{\rmdefault}{\mddefault}{\updefault}{\color[rgb]{0,0,1}$\psi(x)$ continuous }%
}}}}
\put(3720,-1724){\makebox(0,0)[lb]{\smash{{\SetFigFont{8}{9.6}{\rmdefault}{\mddefault}{\updefault}{\color[rgb]{0,0,1}at vertices}%
}}}}
\put(4856,-1598){\makebox(0,0)[lb]{\smash{{\SetFigFont{8}{9.6}{\rmdefault}{\mddefault}{\updefault}{\color[rgb]{0,0,1}$\psi'(x)$ continuous }%
}}}}
\put(5109,-1724){\makebox(0,0)[lb]{\smash{{\SetFigFont{8}{9.6}{\rmdefault}{\mddefault}{\updefault}{\color[rgb]{0,0,1}at vertices}%
}}}}
\put(2616,-1945){\makebox(0,0)[lb]{\smash{{\SetFigFont{8}{9.6}{\rmdefault}{\mddefault}{\updefault}{\color[rgb]{0,0,0}Eq. \eqref{eq:Jean00}}%
}}}}
\put(2616,-2134){\makebox(0,0)[lb]{\smash{{\SetFigFont{8}{9.6}{\rmdefault}{\mddefault}{\updefault}{\color[rgb]{0,0,0}$S\propto\det\mathcal{M}$}%
}}}}
\put(3878,-2071){\makebox(0,0)[lb]{\smash{{\SetFigFont{8}{9.6}{\rmdefault}{\mddefault}{\updefault}{\color[rgb]{0,0,1}$V(x)=0$}%
}}}}
\put(5046,-2071){\makebox(0,0)[lb]{\smash{{\SetFigFont{8}{9.6}{\rmdefault}{\mddefault}{\updefault}{\color[rgb]{0,0,1}$V(x)=0$}%
}}}}
\put(3594,-2418){\makebox(0,0)[lb]{\smash{{\SetFigFont{8}{9.6}{\rmdefault}{\mddefault}{\updefault}{\color[rgb]{0,0,0}Eq. \eqref{eq:LfctCB}}%
}}}}
\put(5267,-2418){\makebox(0,0)[lb]{\smash{{\SetFigFont{8}{9.6}{\rmdefault}{\mddefault}{\updefault}{\color[rgb]{0,0,0}Eq. \eqref{eq:TraceScd}}%
}}}}
\put(5330,-2766){\makebox(0,0)[lb]{\smash{{\SetFigFont{8}{9.6}{\rmdefault}{\mddefault}{\updefault}{\color[rgb]{0,0,1}$\mu_\alpha=0$}%
}}}}
\put(3626,-2766){\makebox(0,0)[lb]{\smash{{\SetFigFont{8}{9.6}{\rmdefault}{\mddefault}{\updefault}{\color[rgb]{0,0,1}$\lambda_\alpha=0$}%
}}}}
\end{picture}%
  \caption{\it Relations between the several expressions derived in
    the article. The dotted green boxes indicate results valid in the
    most general case whereas blue boxes are valid in particular cases.}
  \label{fig:schema}
\end{figure}
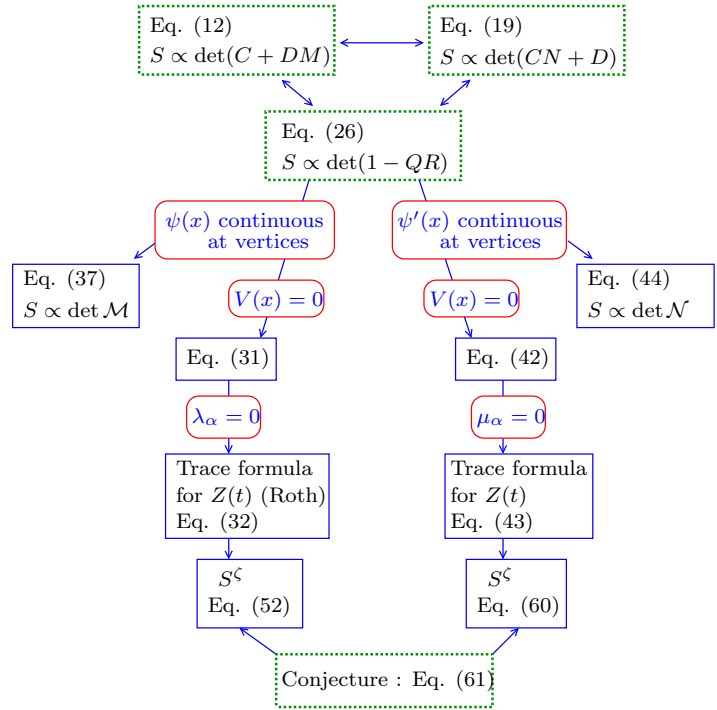

We have summarised the relations between the various expressions of the
spectral determinant given in the article in figure~\ref{fig:schema}.

The work of Desbois \cite{Des01} on the case of general boundary
conditions, in particular equations
(\ref{eq:Sgbc1JD},\ref{eq:DualJean},\ref{eq:Sgbc2JD}), allows in
principle to go continuously from one to the other of these 
two situations by modifying continuously the boundary condition
matrices $C$ and $D$. 
Therefore,
by continuity this leads us to conjecture that the $\zeta$-regularised
spectral determinant of the Schr\"odinger operator for the most general
boundary conditions is given by
\begin{equation}
  \label{eq:conjecture}
  \boxed{
    S^\zeta(\gamma)  = \frac{(-1)^V}{\prod_\alpha{}m_\alpha} 
    \prod_{(a)}\frac{-2}{f'_a(l_a)}\:\det(C+DM) 
    = \frac{\det(\gamma^{-1/4}C-\gamma^{1/4}D)}
           {\prod_\alpha(-m_\alpha)\:\sqrt{\prod_aR_{a,\bar a}}}\,
   \det(\mathbf{1}_{2B}-QR)
  }
\end{equation}
where $\prod_\alpha$ runs over the $V$ vertices, $\prod_{(a)}$ over
the $B$ bonds and $\prod_a$ over the $2B$ arcs.
It would be interesting to provide a direct proof of this formula.

\section*{Acknowledgements}

It is my pleasure to acknowledge useful discussions with Jean Desbois.


\begin{appendix}
  
\section{Bond scattering}
\label{sec:bondscatt}

Let us consider the Schr\"odinger equation on the bond $(a)$ for an
energy $-\gamma=E=+k^2$~:
\begin{equation}
  \left[\gamma-\deriv{^2}{x_a^2}+V_a(x_a)\right]\psi(x_a)=0
  \hspace{0.5cm}\mbox{for }x_a\in(0,l_a)
\end{equation}
We introduce three useful basis of solutions of this differential equation.

\paragraph{Basis 1}
We denote by $\phi_a(x_a)$ the scattering state with an incoming wave
from the left, with
$\phi_a(x_a<0)=\EXP{\I{}kx_a}+r_a\EXP{-\I{}kx_a}$ and
$\phi_a(x_a>l_a)=t_a\EXP{\I{}k(x_a-l_a)}=t_a\EXP{-\I{}kx_{\bar{a}}}$,
where $r_a$, $t_a$ are reflection and transmission amplitudes through
the potential~:
\begin{equation}
  \phi_a(0)=1+r_a  \hspace{0.5cm}\mbox{and}\hspace{0.5cm} \phi_a(l_a)=t_a
  \:.
\end{equation}
The scattering state incoming from the right is naturally denoted by 
$\phi_{\bar{a}}(x_{\bar{a}})$.

A well known sum rule relating integral of the
square modulus of the wave function (i.e. local density of states in
the scattering region) to scattering matrix is the Krein-Friedel sum rule
\cite{Fri52,Kre53,Smi60}. 
Generalisations of this sum rule in the context of metric graphs have
been obtained in Refs.~\cite{Tex02,TexBut03}. 
For $E=k^2=-\gamma>0$, let us change notation and introduce
$\Psi^{(L)}(x)=\frac1{\sqrt{4\pi k}}\phi_a(x)$ and 
$\Psi^{(R)}(x)=\frac1{\sqrt{4\pi k}}\phi_{\bar{a}}(l_a-x)$, the
left and right stationary scattering states, respectively.
We can deduce from the local version of the Krein-Friedel sum rule
demonstrated in Ref.~\cite{TexBut03} for graphs that
\begin{equation}
   \int_0^{l_a} \D x\, \big[\Psi^{(\alpha)}(x)\big]^* \Psi^{(\beta)}(x)
   = 
  \frac1{2\I\pi}
  \left(
    \mathscr{S}^\dagger\deriv{\mathscr{S}}{E}
    +\frac{\mathscr{S}-\mathscr{S}^\dagger}{4E}
  \right)_{\alpha\beta}
  \hspace{0.5cm}
  \mbox{with } \alpha,\,\beta\in\{L,\,R\}
\end{equation}
where $\mathscr{S}$ is the $2\times2$ scattering matrix describing the
scattering by the potential on the bond
\begin{equation}
    \mathscr{S} =
    \begin{pmatrix}
       r_a & t_{\bar a} \\ t_a & r_{\bar a}
    \end{pmatrix}
  \:.
\end{equation}
We obtain
\begin{align}
    \int_0^{l_a} \D x\, |\phi_a(x)|^2 &= 
  -2\I\sqrt{E} 
  \left( 
    r_a^*\deriv{r_a}{E} + t_a^*\deriv{t_a}{E} + \frac{r_a-r_a^*}{4E}  
  \right)\\
    \int_0^{l_a} \D x\, \phi_a(x)^*\phi_{\bar a}(l_a-x) &= 
  -2\I\sqrt{E} 
  \left( 
    r_a^*\deriv{t_{\bar a}}{E} + t_a^*\deriv{r_{\bar a}}{E} 
   + \frac{t_{\bar a}-t_a^*}{4E}  
  \right)
  \:.
\end{align}

\paragraph{Basis 2}
Let us introduce the solution $f_a(x_a)$ satisfying the boundary conditions
\begin{equation}
  f_a(0)=1  \hspace{0.5cm}\mbox{and}\hspace{0.5cm} f_a(l_a)=0
  \:.
\end{equation}
It follows that $f_a(x)$ is a real function for $\gamma\in\mathbb{R}$.
For example, when $V(x)=0$, the function reads
$f_a(x)=\frac{\sinh\sqrt\gamma(l_a-x)}{\sinh\sqrt\gamma{}l_a}$.
Another independent
solution of this differential equation is naturally denoted
$f_{\bar{a}}(l_a-x_a)=f_{\bar{a}}(x_{\bar{a}})$ 
and take the values $0$ for $x_a=0$ and $1$ for
$x_a=l_a$ (do not confuse the two functions
$f_a$ and $f_{\bar{a}}$ with the components of a scalar function).
The Wronskian of the two solutions is
\begin{equation}
  \label{eq:Wronskif}
  {W}_a 
  = \mathcal{W}[f_a(x),f_{\bar{a}}(l_a-x)]=-f'_a(l_a)=-f'_{\bar{a}}(l_a)
  \:.
\end{equation}
Sum rules can be obtained as follows~\cite{Des00}~:
we remark that $\partial_\gamma{}f_a(x)$ is solution of the
differential equation
$\big[\gamma-\deriv{^2}{x^2}+V_a(x)\big]\partial_\gamma{}f_a(x)=-f_a(x)$
for the boundary conditions
$\partial_\gamma{}f_a(0)=\partial_\gamma{}f_a(l_a)=0$. 
Integration straightforwardly gives
\begin{equation}
  \partial_\gamma{}f_a(x) = 
  -\frac1{{W}_a}
  \left[
     f_a(x)\int_0^x\D{}x'\,f_a(x')f_{\bar{a}}(l_a-x')
    +f_{\bar{a}}(l_a-x)\int_x^{l_a}\D{}x'\,f_a^2(x')
  \right]
  \:.\label{eq:1}
\end{equation}
We deduce
\begin{align}
  \label{eq:RelDes00a}
  \int_0^{l_a}\D{}x\,f_a(x)^2 &= -\partial_\gamma{}f'_a(0)\\
  \label{eq:RelDes00b}
  \int_0^{l_a}\D{}x\,f_a(x)f_{\bar{a}}(l_a-x)
  &= \partial_\gamma{}f'_a(l_a)
  \:.
\end{align}
They replace the Krein-Friedel sum rule like formulae given above for
scattering states.

One can establish the relation between the two basis of solutions
$\{\phi_a(x_a),\,\phi_{\bar{a}}(x_{\bar{a}})\}$
and 
$\{f_a(x_a),\,f_{\bar{a}}(x_{\bar{a}})\}$
\cite{TexMon01}~:
\begin{equation}
  f_a(x_a) =
  \frac{(1+r_{\bar{a}})\phi_a(x_a)-t_a\phi_{\bar{a}}(x_{\bar{a}})}
       {(1+r_a)(1+r_{\bar{a}})-t_at_{\bar{a}}}
  \:.  
\end{equation}
The two following relations follow
\begin{align}
  \label{eq:fpa}
  f'_a(0) &= -\sqrt\gamma\,
 \frac{(1-r_a)(1+r_{\bar{a}})+t_at_{\bar{a}}}{(1+r_a)(1+r_{\bar{a}})-t_at_{\bar{a}}} 
  \\
  \label{eq:fpab}
  f'_a(l_a) &= -\sqrt\gamma\,
 \frac{2\,t_a}{(1+r_a)(1+r_{\bar{a}})-t_at_{\bar{a}}}
\end{align}
where we have performed some analytic continuation to negative
energies $\gamma=-k^2-\I0^+$.
Let us introduce the $2\times2$ block in the matrix $M$
related to the two arcs $a$ and $\bar{a}$~:
\begin{equation}
  M_{(a)} =
  \begin{pmatrix}
    f'_a(0) & -f'_{\bar{a}}(l_a)\,\EXP{\I\theta_{\bar a}} \\ 
    -f'_a(l_a)\,\EXP{\I\theta_{a}} & f'_{\bar{a}}(0)
  \end{pmatrix}
  \:.
\end{equation}
The relations (\ref{eq:fpa},\ref{eq:fpab}) with \eqref{eq:RelRN}
immediately shows that the $2\times2$ block in the matrix $R$
related to the two arcs $a$ and $\bar{a}$ indeed encodes the reflection
and transmission amplitudes~:
\begin{equation}
  R_{(a)} =
  \begin{pmatrix}
    r_a & t_{\bar{a}}\,\EXP{\I\theta_{\bar a}} \\ 
    t_a\,\EXP{\I\theta_{a}} & r_{\bar{a}}
  \end{pmatrix}
  \hspace{0.5cm}\mbox{that is}\hspace{0.5cm}
  R_{a,b} = \delta_{a,b}\,r_a 
  + \delta_{a,\bar b}\,t_{\bar a}\,\EXP{\I\theta_{\bar a}}
  \:,
\end{equation}
Hence demonstrating that $R$ is the ``bond scattering matrix'' (or
more precisely its analytic conitnuation to energy $E\to-\gamma$).
We obtain the useful relation
\begin{equation}
  \label{eq:useful6}
  -f'_a(l_a)\,\det(\mathbf{1}_{2}+R_{(a)}) = 2\sqrt\gamma\,t_a
\end{equation}
(we recall that $t_a=t_{\bar{a}}$ follows from the one-dimensional
character of the wire~: it can be demonstrated by computing the
Wronski determinant of the two scattering states at the two edges of
the interval).

\paragraph{Basis 3}
Another useful set of solutions are the functions $g_a(x_a)$ and
$g_{\bar{a}}(x_{\bar{a}})$  whose {\it derivative} take values $1$ and
$0$ at the two ends of the interval~:
\begin{equation}
  g'_a(0)=1  \hspace{0.5cm}\mbox{and}\hspace{0.5cm} g'_a(l_a)=0
  \:.
\end{equation}
For example, when $V(x)=0$, the function reads
$g_a(x)=-\frac{\cosh\sqrt\gamma(l_a-x)}{\sqrt\gamma\sinh\sqrt\gamma{}l_a}$.
The Wronski determinant is given by 
\begin{equation}
  {W}^g_a 
  = \mathcal{W}[g_a(x),g_{\bar{a}}(l_a-x)]=-g_a(l_a)=-g_{\bar{a}}(l_a)
  \:.
\end{equation}
Following the method of Ref.~\cite{Des00} and recalled above one can
construct the derivative of the function with respect to the spectral parameter
$\partial_\gamma{}g_a(x)=-\frac1{{W}^g_a}\big[g_a(x)\int_0^x\D{}x'\,g_a(x')g_{\bar{a}}(l_a-x')+g_{\bar{a}}(l_a-x)\int_x^{l_a}\D{}x'\,g_a^2(x')\big]$.
Therefore
\begin{align}
  \int_0^{l_a}\D{}x\,g_a(x)^2 &= \partial_\gamma{}g_a(0)\\
  \int_0^{l_a}\D{}x\,g_a(x)g_{\bar{a}}(l_a-x) &= \partial_\gamma{}g_a(l_a)
\end{align}
It is also interesting to relate this basis of solutions to the two
other basis~:
\begin{equation}
  g_a(x_a) = -\frac{1}{\sqrt\gamma}
  \frac{(1-r_{\bar{a}})\phi_a(x_a)+t_a\phi_{\bar{a}}(x_{\bar{a}})}
       {(1-r_a)(1-r_{\bar{a}})-t_at_{\bar{a}}}
\end{equation}
from which we obtain
\begin{align}
  g_a(0) &= -\frac1{\sqrt\gamma}\,
 \frac{(1+r_a)(1-r_{\bar{a}})+t_at_{\bar{a}}}{(1-r_a)(1-r_{\bar{a}})-t_at_{\bar{a}}} 
  \\
  g_a(l_a) &= -\frac1{\sqrt\gamma}\,
 \frac{2\,t_a}{(1-r_a)(1-r_{\bar{a}})-t_at_{\bar{a}}}
  \:.
\end{align}
We also easily demonstrate that 
\begin{equation}
  g_a(x_a) =
  \frac{f'_{\bar{a}}(0)f_a(x_a)+f'_a(l_a)f_{\bar{a}}(x_{\bar{a}})}
       {f'_a(0)f'_{\bar{a}}(0)-f'_a(l_a)f'_{\bar{a}}(l_a)}   
\end{equation}
from which 
\begin{align}
  g_a(0) &= 
    \frac{f'_{\bar{a}}(0)}{f'_a(0)f'_{\bar{a}}(0)-f'_a(l_a)f'_{\bar{a}}(l_a)}\\
  g_a(l_a) &=
    \frac{f'_{\bar{a}}(l_a)}{f'_a(0)f'_{\bar{a}}(0)-f'_a(l_a)f'_{\bar{a}}(l_a)}
\end{align}
The two matrices introduced above are hence simply related by
\begin{equation}
  N_{(a)} =
  \begin{pmatrix}
    g_a(0) & g_{\bar{a}}(l_a)\,\EXP{\I\theta_{\bar a}} \\ 
    g_a(l_a)\,\EXP{\I\theta_{a}} & g_{\bar{a}}(0)
  \end{pmatrix}
  = M_{(a)}^{-1}
  \:.
\end{equation}
Note also that 
$\det M_{(a)}=f'_a(l_a)/g_a(l_a)$.

\section{$\zeta$-functions for combinatorial graphs}

The study of $\zeta$-functions (also denoted $L$-functions in the
mathematical literature) and  trace formulae in combinatorial graphs
has attracted a lot of interest
\cite{Iha66,Has89,Has90,Bas92,StaTer96,Bar99} (they found recently
some application in the context of quantum chaos
\cite{OreGodSmi09,OreSmi10})~; a brief recent review on 
mathematical aspects may be found in the introduction of Ref.~\cite{MizSat05}.
We show in this appendix that the general trace formula 
for combinatorial graphs (the Bartholdi's formula~\cite{Bar99}
generalising Bass~\cite{Bas92} and Ihara~\cite{Iha66} trace formulae)
may be deduced from the trace formula obtained in 
Ref.~\cite{AkkComDesMonTex00} for metric graphs. This latter is itself
a particular case of the general trace formula obtained by Desbois
\cite{Des01} recalled in section~\ref{sec:Desbois}.

We consider the trace formula for the Laplace operator acting on
functions continuous at the vertices.
Our starting point is the equality, Eqs.~\eqref{eq:Sgbc2JD} and
\eqref{eq:Jean00}, 
\begin{equation}
  \label{eq:88}
  (-1)^V\frac{\det(C-\sqrt\gamma\, D)}{\det(\mathbf{1}_{2B}+R)}\,
  \det(\mathbf{1}_{2B}-QR)
  =\det\mathcal{M}  
\end{equation}
We recall that in this case
$(-1)^V\det(C-\sqrt\gamma\,D)=\prod_\alpha(\lambda_\alpha+m_\alpha\sqrt\gamma)$. 

Note that \eqref{eq:88} is related to the spectral determinant of the
metric graph when boundary condition matrices $C$ and $D$ are $\gamma$
independent. However, having established the relation between
the vertex determinant and the arc determinant, Eq.~\eqref{eq:88} also holds
when the parameters $\lambda_\alpha$ depend 
on $\gamma$ (the case considered below).
In this case, Eq.~\eqref{eq:88}
is however not anymore related to the spectral determinant of the
metric graph. 

Let us consider the case with no potential $V(x)=0$ and set all the
lengths of the wires equal~: $l_a=l\:\forall\:a$. 
For convenience, we introduce the notation
\begin{equation}
    \EXP{-\sqrt\gamma l} = w u
  \:,
\end{equation}
where $u$ and $w$ are two real or complex numbers.
The bond scattering matrix elements are
$R_{ij}=wu\,\delta_{i,\bar j}$ and therefore 
$\det(\mathbf{1}_{2B}+R)=[1-(uw)^2]^B$.
We consider the case where the boundary condition parameters
$\lambda_\alpha$ are related to the valencies of the vertex by the
relation 
\begin{align}
  m_\alpha+ \frac{\lambda_\alpha}{\sqrt\gamma} = 2w
  \:.
\end{align}
The matrix $QR$ has elements~:
$(QR)_{ij}=u$ if the arc $j$ terminates at the vertex from which issues the
arc $i$ (with $\bar j\neq i$)~;
$(QR)_{i\,\bar i}=u(1-w)$~; all other matrix elements are zero. 
We write (notations reminiscent of those of  Refs.~\cite{Bar99,MizSat05})
\begin{equation}
QR=u(\mathscr{B}-w\mathscr{J})
\end{equation}
where 
$\mathscr{B}_{ij}=1$ if end$(j)=$begining$(i)$, $\mathscr{B}_{ij}=0$ otherwise~; the
matrix $\mathscr{J}$ couples reversed arcs $\mathscr{J}_{ij}=\delta_{i,\bar j}$.

On the other hand, from \eqref{eq:PasMon} we find
\begin{equation}
    \mathcal{M}_{\alpha\beta}  = \frac{2\sqrt\gamma}{1-(uw)^2}
  \left\{
    \delta_{\alpha\beta} \left[w+(m_\alpha-w)(uw)^2\right]
    - u w\, a_{\alpha\beta}
  \right\}
\end{equation}

Using that $(-1)^V\det(C-\sqrt\gamma\,D)=(2w\sqrt\gamma)^V$ we
deduce the Bartoldi's formula for the zeta function
\cite{Bar99,Des01,MizSat05,OreGodSmi09}
\begin{align}
  \label{eq:Bartholdi}
  &Z(u,w)^{-1}=\prod_{\widetilde{\mathcal{C}}}
  \left( 
    1 - (1-w)^{n_R(\widetilde{\mathcal{C}})}\,u^{\ell(\widetilde{\mathcal{C}})}
  \right) \\
  &= \det(\mathbf{1}_{2B}-u(\mathscr{B}-w\mathscr{J}))
  = (1-(uw)^2)^{B-V} \det(\mathbf{1}_{V}(1-(uw)^2) - u \mathcal{A} + wu^2\mathcal{Y})  
  \:,
\end{align}
where $\ell(\mathcal{C})$ is the length of the orbit
(number of wires visited) and $n_R(\mathcal{C})$ 
the number of reflections of the orbit on vertices.
$\mathcal{A}$ denotes the adjacency matrix with matrix elements
$a_{\alpha\beta}$. 
The matrix $\mathcal{Y}$ is the diagonal matrix gathering the valencies~:
$\mathcal{Y}_{\alpha\beta}=m_\alpha\delta_{\alpha\beta}$. 
This formula was used in Ref.~\cite{Des01,ComDesTex05} as a generating
function for counting of the orbits with finite number of reflections. 
It has also found some interesting application in the context of
quantum chaos in combinatorial graphs \cite{OreGodSmi09,OreSmi10}.

Note that combinatorial graphs are related to metric graphs when all
lengths are equal and for permutation invariant boundary conditions at
the vertices. It follows that only two weights can be attributed to an
orbit passing at a vertex ($u$ for no reflection and $u(1-w)$ for a
reflection) and therefore \eqref{eq:Bartholdi} is the most general
trace formula for combinatorial graphs (up to the straightforward
addition of magnetic fluxes in matrices $\mathcal{A}$ and $\mathscr{B}$).
As a consequence all the trace formulae obtained for metric graphs
with permutation invariant boundary conditions at
the vertices (we have consider the particular case of continous
boundary conditions here) would lead to the Bartholdi formula when
lengths are taken to be equal and potential vanishes. 

For $w=1$ we recover the Bass formula \cite{Bas92,StaTer96} (this was also
discussed in Refs.~\cite{Des01,ComDesTex05}) 
\begin{align}
  \label{eq:Bartholdi}
  Z(u,1)^{-1}&=\prod_{\widetilde{\mathcal{C}}_B}
  \left(1-u^{\ell(\widetilde{\mathcal{C}}_B)}\right)\\
  &= \det(\mathbf{1}_{2B}-u(\mathscr{B}-\mathscr{J}))
  = (1-u^2)^{B-V} \det(\mathbf{1}_{V} - u \mathcal{A} + u^2(\mathcal{Y}-\mathbf{1}_{V}))  
\end{align}
for the $\zeta$-function over the backtrack-less orbits
$\widetilde{\mathcal{C}}_B$ (orbits with no reflection).
The Bass formula generalises the Ihara formula \cite{Iha66}
demonstrated for regular graphs.

\end{appendix}


\end{document}